\documentclass[fleqn,usenatbib]{mnras}

\usepackage{newtxtext,newtxmath}

\usepackage[T1]{fontenc}
\usepackage{ae,aecompl}

\usepackage{amssymb}
\usepackage{verbatim}
\usepackage{graphicx}
\usepackage{bigfoot}
\usepackage{bm}
\newcommand{\rmd}{{\rm d}}

\newcommand{\sgb}{\sigma_{\rm b}}
\newcommand{\sgc}{\sigma_{\rm c}}
\newcommand{\sgo}{\sigma_{\rm o}}
\newcommand{\sgq}{\sigma_{\rm q}}
\newcommand{\sgr}{\sigma_{\rm r}}
\newcommand{\sgs}{\sigma_{\rm s}}
\newcommand{\sgt}{\sigma_{\rm t}}

\newcommand{\sgbsq}{\sigma_{\rm b}^2}
\newcommand{\sgcsq}{\sigma_{\rm c}^2}
\newcommand{\sgosq}{\sigma_{\rm o}^2}
\newcommand{\sgqsq}{\sigma_{\rm q}^2}
\newcommand{\sgrsq}{\sigma_{\rm r}^2}
\newcommand{\sgssq}{\sigma_{\rm s}^2}
\newcommand{\sgtsq}{\sigma_{\rm t}^2}

\newcommand{\DQ}{\Delta Q}
\newcommand{\DN}{\Delta N}
\newcommand{\dT}{\delta T}
\let\ni=\noindent

\title[ISW in $\Lambda$CDM or something else?]
      {ISW in $\Lambda$CDM or something else?}

\author[A. M. So\l tan]{A. M. So\l tan\thanks{E-mail:
soltan@camk.edu.pl}\\
Nicolaus Copernicus Astronomical Centre, Polish Academy of Science,
Bartycka 18, 00-716 Warsaw, Poland }

\begin{document} 
\date{Accepted \hspace{15mm}. Received  \hspace{15mm}; in original form }

\pagerange{\pageref{firstpage}--\pageref{lastpage}} \pubyear{}

\maketitle

\label{firstpage}

\begin{abstract} We investigate a correlation between the {\it Planck's} CMB
temperature map and statistics based on the space density of quasars in the
SDSS catalogue. It is shown that the amplitude of the positive correlation
imposes a lower limit on the amplitude of the Integrated Sachs-Wolfe (ISW)
effect independent of the quasar bias factor. Implications of this constraint
for the ISW effect in the $\Lambda$CDM model are examined.  Strength of the
correlation indicates that the rms of temperature fluctuations associated with
the quasars distributed between $1500$ and $3000$\,Mpc likely exceeds
$11-12\,\mu$K. The signal seems to be related to an overall space distribution
of quasars rather than to a few exceptionally dominant structures like
supervoids. Although, the present estimates are subject to sizable
uncertainties, the signal apparently exceeds the model predictions of the ISW
effect for the standard $\Lambda$CDM cosmology.  This conclusion is consistent
with several other investigations that also claim some disparity between the
observed ISW signal and the theoretical predictions.

\end{abstract}


\section{Introduction}

Distribution of the cosmic microwave background (CMB) temperature in the
celestial sphere is determined mostly by local matter parameters in the
recombination era at a redshift of $z \approx 1100$. CMB photons along their
path to the observer are subject to various interactions (e.g.
\citealt{dodelson03}). In particular, variations of gravitational potential
induced by the large-scale matter density fluctuations affect photon energies
that introduce additional temperature fluctuations, known as Rees-Sciama effect
\citep{rees68}. In the Universe with cosmological constant amplitude of the
effect is dominated by the linear term of density fluctuations, and the
phenomenon is called Integrated Sachs-Wolfe (ISW) effect \citep{sachs67}.

Measurements of the CMB temperature deviation at the position of large-scale
matter agglomerations would be the straightforward  way to investigate the ISW
effect. However, this attitude encounters difficulties, because the ISW signal
generated by a single supervoid or supercluster is substantially weaker than
CMB fluctuations originating at the last scattering surface. To improve
signal-to-noise ratio one needs extensive catalogues of extragalactic
objects (galaxies, quasars, radio sources) that allow for selection of a number
of superclusters and voids. Fluctuations of gravitational potential associated
with such structures presumably generate the strongest ISW signal. Stacking the
CMB temperature maps centered at superstructers allows for effective assessment
of the ISW signal.  If the analysis is not confined  to prominent structures,
but encompasses the whole distribution of objects over the a large area of
the sky, the ISW effect is investigated using correlation analysis
between both distributions, i.e. temperature and matter tracers. In this case
several statistical tools to examine the ISW effect were applied of which two
are most commonly used, namely: the angular cross-correlation function (CCF) in
real space, and cross-angular power spectrum (CAPS) correlations in the harmonic
space. i.e. the correlation between the spherical harmonic coefficients.

Clearly, all the methods should provide comparable results, and in fact in most
cases statistical uncertainties are sufficiently large to ensure consistency
between different ISW measurements. However, all these methods also suffer from
systematic errors that apparently lead to systematic differences between
stacking and correlation analyses. The investigations based on stacking the CMB
temperature maps detect generally stronger ISW signal than those based on the
CCF or CAPS. Amplitude of the ISW effect is quantified typically in two ways.
If the stacking method is applied, the average signal generated by
superclusters and voids is given in $\mu$K, while the correlation analyses
usually provide relative strength of the effect normalized to the amplitude
expected for the $\Lambda$CDM model. In this case, theoretical predictions are
assessed using cosmological simulations, or the power spectrum of matter
fluctuations in the local Universe derived for the assumed cosmological
parameters. 

A positive detection of the ISW effect has been reported in a large number of
investigations based on extensive CMB data gathered in WMAP and {\it Planck}
missions, and several surveys of discrete objects. A logical next step is to
examine if the strength of the ISW signal is consistent with that expected for
the standard $\Lambda$CDM cosmology. Such analysis requires precise assessment
of the observed signal as well as accurate model calculations.  However, both
those aspects are still subject to statistical and systematic errors. In
effect, a question whether the observed amplitude of the ISW effect
conforms to the $\Lambda$CDM is still debatable.

Surprisingly strong ISW signal was reported by  \citet[therein references to
some earlier works]{granett08}. They used a large sample of luminous red
galaxies from the SDSS \citep{adelman08} populating a volume of $\sim
5\,h^{-3}$ Gpc$^3$ in the redshift range $0.4 < z < 0.75$. Most prominent
supervoids and superclusters, i.e. extended low and high density
areas, were carefully selected using dedicated algorithm. Then, the CMB
temperature distribution in maps from the WMAP $5$ year survey
\citep{bennett03,hinshaw09} was investigated.  Using various sets of voids and
clusters, as well as different filters to assess the temperature signal they
found systematic difference between the average temperatures of clusters and
voids. In the most significant case of $50$ clusters and $50$ voids, the
average temperature deviation amounts to $7.9\,\mu$K for clusters and
$-11.3\,\mu$K for voids. The amplitudes represent above $4\,\sigma$ {\it a
posteriori} detection that \citet{granett08} attributed to the ISW effect.

In a series of follow-up papers several other groups obtain results similar or
at least comparable to the original \citet{granett08} signal. Here we recall
several investigations most relevant for the present analysis. We begin with
reports based on stacking technique.  \citet{ilic13} basically confirmed
Granett's et al. results, although their analysis using the new void catalogue
by \citet{sutter12} gave weaker ISW signature and revealed some differences in
the relationship between the extent of the ISW signal and the void size. Using
Jubilee simulations data several authors \citep[e.g.][]{hotchkiss15} show that
superstructures in $\Lambda$CDM generate the ISW signal substantially
smaller than \citet{granett08} measurement. Consequently, amplitudes order of
magnitude larger must `arise from something other than an ISW effect in a
$\Lambda$CDM universe'.  Similar conclusion was reached by \citet{kovacs17}.
Using {\it Planck's} SMICA map \citep{planckXI} and the DES catalogue
\citep{flaugher15,darkenergy16} the authors got excessive cold and hot
imprints in voids and superclusters, respectively, although of low statistical
significance. \citet{nadathur16} identified  $2445$ voids and $29866$
superclusters in the galaxy CMASS sample of the SDSS-III BOSS DR 12
\citep{alam15a} and stacked them in bins according to their gravitational
potential strength. Using the {\it Planck's} CMB data, they determined the
average temperature deviation for each bin separately.  With a calibration of
the ISW signal based on the Big MD $N$-body simulations \citep{klypin16}
\citet{nadathur16} measure the ISW amplitude at $1.64 \pm 0.53$ relative to
the $\Lambda$CDM expectation, in agreement with the model.  Nevertheless, in
the context of the present investigation we note that the best fit to the data
is $1.2\,\sigma$ above the $\Lambda$CDM level.  Furthermore, the most
prominent voids produce the temperature drop above $10\,\mu$K, and the largest
superclusters temperature excess above $5\mu$K, admittedly with large
uncertainties. According to \citet{nadathur16}. these estimates exceed by a
factor of two the figures predicted for the $\Lambda$CDM model.
\citet{planckXIX} and \citet{planckXXI} investigated the ISW effect applying
several statistics, CCF, CAPS and stacking among others. Both correlation
analyses for several galaxy and AGN samples find the ISW amplitude fully
consistent in statistical sense with the $\Lambda$CDM universe. Although,
systematic differences between the samples weaken to some extent this
conclusion. On the other hand, their stacking analysis confirmed high figures
obtained by \citet{granett08}.  Also for the largest voids in the
\citet{sutter12} catalogue \citet{planckXIX} reports the signal above what is
expected from simulations, although weaker than in $50$ Granett's et al.
voids.

Numerous investigations based on correlations create a different
picture.  \citet{raccanelli08} detected the ISW signal in the WMAP $3$ year
data \citep{hinshaw07} generated by the NRAO VLA Survey \citep{condon98} and
concluded that its amplitude is fully consistent with the predictions of the
standard $\Lambda$CDM cosmology. Using correlations in harmonic space between
the AllWISE catalogue of the WISE survey \citep{wright10} and the WMAP $9$
year data \citep{bennett13} \citet{shajib16} also report apparent consistency
of the ISW effect with the $\Lambda$CDM model.
Similar conclusion was reached also by
\citet{granett09}. \citet{granett15} reinvestigated a question of consistency
between the amplitude of the detected ISW signal and the signal predicted for
the $\Lambda$CDM model. They noticed that their results differ from those by
\citet{granett09} and conclude that the amplitude ratio depends on the number
density, redshift distribution and redshift uncertainties of the matter
tracers. Consequently, they restrain themselves from drawing conclusion on that
point.  A related question was discussed by \citet{ho08}. These authors  test
feasibility to constrain parameters of the $\Lambda$CDM model by means of the
ISW effect, using correlations of $3$ year WMAP data with a number of
catalogues of discrete sources.

\citet{hernandez13} analyze a correlation between the WMAP data and the
catalogue of clusters and voids from the original Granett's et al paper.
They find that the detected signal is incompatible with standard $\Lambda$CDM
cosmology.  However, statistical significance of the discrepancy is lowered
from $\sim\!4\,\sigma$ claimed by \citet{granett08} to $\sim\!2.2\,\sigma$, if
one takes into account a narrow angular range of the effect. A mild excess
signal with respect to the expectations from the $\Lambda$CDM model, based on
the correlations between the WMAP and several surveys report
\citet{giannantonio12}. In a recent paper \citet{stoelzner18} derive
constraints on the ISW effect through correlations of the {\it Planck's}
temperature maps with several galaxy, AGN and radio source catalogues. Their
scrupulous statistical analysis confirmed the ISW effect at significance level
reaching $5\,\sigma$, with the amplitude normalized to the $\Lambda$CDM model
`above $1$ at around $1\,\sigma$ or a bit more'.

Recently \citet{kovacs18} compared the ISW signal expected from the
structures identified as {\it supervoids} in the Jubilee simulations with
that observed for stacked voids in the BOSS DR12 catalogue. He reports a factor
of up to $9$ excess signal in the real data with possible $20$ percent
differences in the radial profiles of simulated and actually observed voids.

One can expect that assorted statistical techniques applied to various
observational data should provide consistent results of the ISW effect.  In
fact, the very correlation between distributions of discrete objects and
the CMB temperature has been established using both techniques, i.e.
correlating preselected supersctructures with the CMB, as well as using large
sky areas that are representative for the overall matter distribution.
However, results  based on the wide area correlations seem to be compatible
with the ISW effect predicted in the $\Lambda$CDM cosmology, while
investigations that utilize stacking of large supervoids and
superclusters provide estimates substantially above the theoretical
predictions.

The measured correlation amplitudes are determined directly from observations
On the other hand, a question whether these amplitudes are consistent with the
predicted strength of the ISW effect in the $\Lambda$CDM cosmology involves a
series of assumptions that are strongly model dependent.  Thus, it is likely
that discordant conclusion on this point result from numerous systematics. The
amplitude of the ISW effect in the $\Lambda$CDM cosmology involves detailed
modelling of various physical processes. The ISW effect is generated by large
scale matter agglomerations. Their properties are adequately described by
linear theory, while the discrete observable objects are formed in highly
nonlinear processes. To test if the ISW effect is consistent with the
$\Lambda$CDM cosmology one needs to determine relationships between statistical
characteristics of large invisible structures and observable discrete objects.
In principle, cosmological simulations that reproduce observable distributions
of various classes of discrete sources would ultimately provide sought
relationship. However, the complexity of physical processes describing the
evolution of baryonic matter that eventually build up luminous, discrete
objects to some extent restricts the present modelling to phenomenological
methods. Therefore, quantitative investigation of the ISW effect in the
simulations should be treated with caution.

In the present paper we test a method that is based solely on statistical
properties of the CMB fluctuations and the quasar distribution. Therefore our
results are not subject to systematic uncertainties induced by modeling of
physical processes that involve dark matter and discrete objects. At the same
time the present estimates can be directly compared with the predictions of the
ISW effect obtained in cosmological models. We do not try to reconstruct a 3D
map of the gravitational potential. Rather than to find a one-to-one
correspondence between the discrete objects and the potential, we define an
observable parameter proportional to the local concentration of objects. This
parameter is expected to correlate in statistical sense with the potential. In
effect, we resign from the quasar bias factor and reduce influence of the
redshift distribution on final results.  Statistics based solely on the
quasar distribution without relation to the bias factor does not give definite
estimates of the ISW signal. Nevertheless, it provides potentially
restrictive  lower limits for the contribution of the ISW effect. These
statistics are described in details in Sec.~\ref{sec:qso_phi} and the Appendix.

Since the ISW signal is proportional to a net change of the gravitational
potential during photon travel time, the dominant contribution to the amplitude
of the effect comes from large-linear-scale fluctuations of the total matter
density. Because the large volumes are involved, only the intrinsically
luminous objects, detectable over large distances, are suitable for the
analysis.  In most of the previous investigations bright galaxies have been
used as matter tracers. In this paper we use quasars for two reasons. First,
quasar samples cover usually huge volumes. Second, criteria for quasar
selection are different than those for galaxies. Therefore, from the data
acquisition point of view, quasars provide information on the large-scale
matter distribution that is independent from the galaxy data.
Concentration of quasars is substantially lower than the galaxy space density,
and the average distance between neighbouring quasars is much larger than
between the SDSS galaxies. However, clustering properties of quasars and
galaxies are not distinctly different even at small scales \citep{ross09}.
Thus, at scales of several hundreds Mpc the quasars are equally adequate
tracers of the luminous matter, and may be used to measure the total matter
density similarly to galaxies (see below).  In those cases where the present
paper covers the same area as some other work, in particular the
\citet{granett08},  both types of objects are expected to provide comparable
results.

All distances and linear dimensions are expressed in co-moving coordinates.  To
convert redshifts to the co-moving distances, we use the flat cosmological
model with $H_{\rm o} = 70$\,km\,s$^{-1}$Mpc$^{-1}$, $\Omega_{\rm m} = 0.30$
and $\Omega_{\Lambda} = 0.70$. We focus our study on two distance areas:
`near' $\equiv 1500 - 3000$\,Mpc, and `distant' $\equiv 3000 - 4500$\,Mpc, what
correspond approx. to redshifts $0.4 - 0.9$ and $0.9 - 1.6$.

The paper is organized as follows. In the next section the observational
material used in the investigation is described. It includes a short
description of the quasar sample and basic information on the {\it Planck's}
CMB data.  In Sec.~\ref{sec:qso_isw} standard formulae of the ISW effect are
recalled and statistics used to process the quasar data is defined.  In
Sec.~\ref{sec:results} the amplitude of the correlations between the CMB
temperature map and the quasar statistics are assessed.  The lower limits of
the ISW effect generated by the matter distribution at distance bins
$1500-3000$ and $3000-4500$\,Mpc are obtained. The results are discussed in
Sec.~\ref{sec:discussion}. In the Appendix we derive some formulae relevant for
the present investigation. In particular those involved in the linear
correlation analysis.


\section{The data}
 \label{sec:data}

\subsection{SDSS quasar catalogue}
\begin{figure}
   \includegraphics[width=1.00\linewidth]{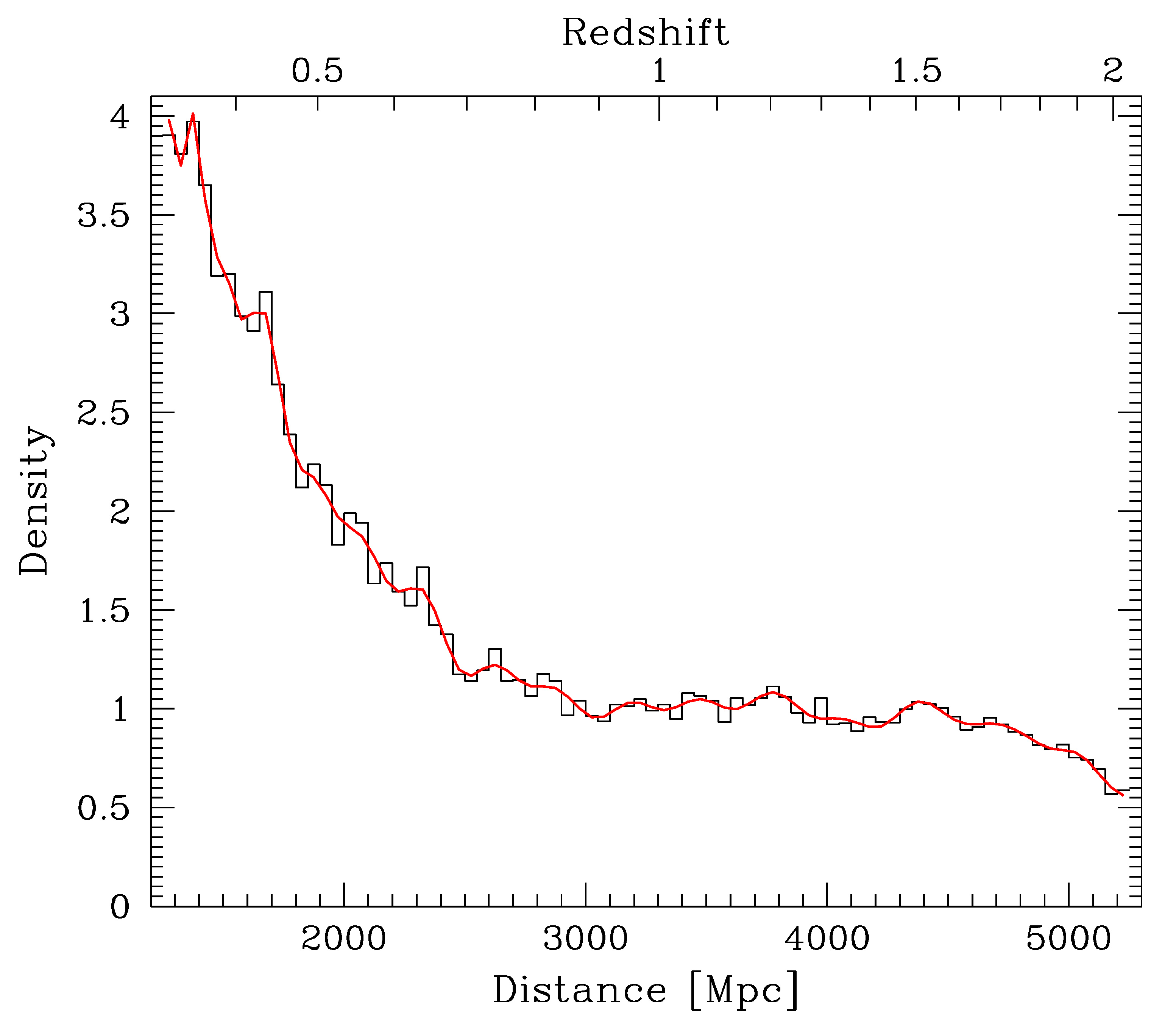}
   \caption{Distance distribution of quasar space density (arbitrary units)
   of the SDSS quasars  brighter than $z = 19.50$.
   Superimposed curve is a Legendre polynomial fit of degree $n=50$}.
   \label{fig:distance_fit}
\end{figure}
   
\begin{figure}
   \includegraphics[width=1.00\linewidth]{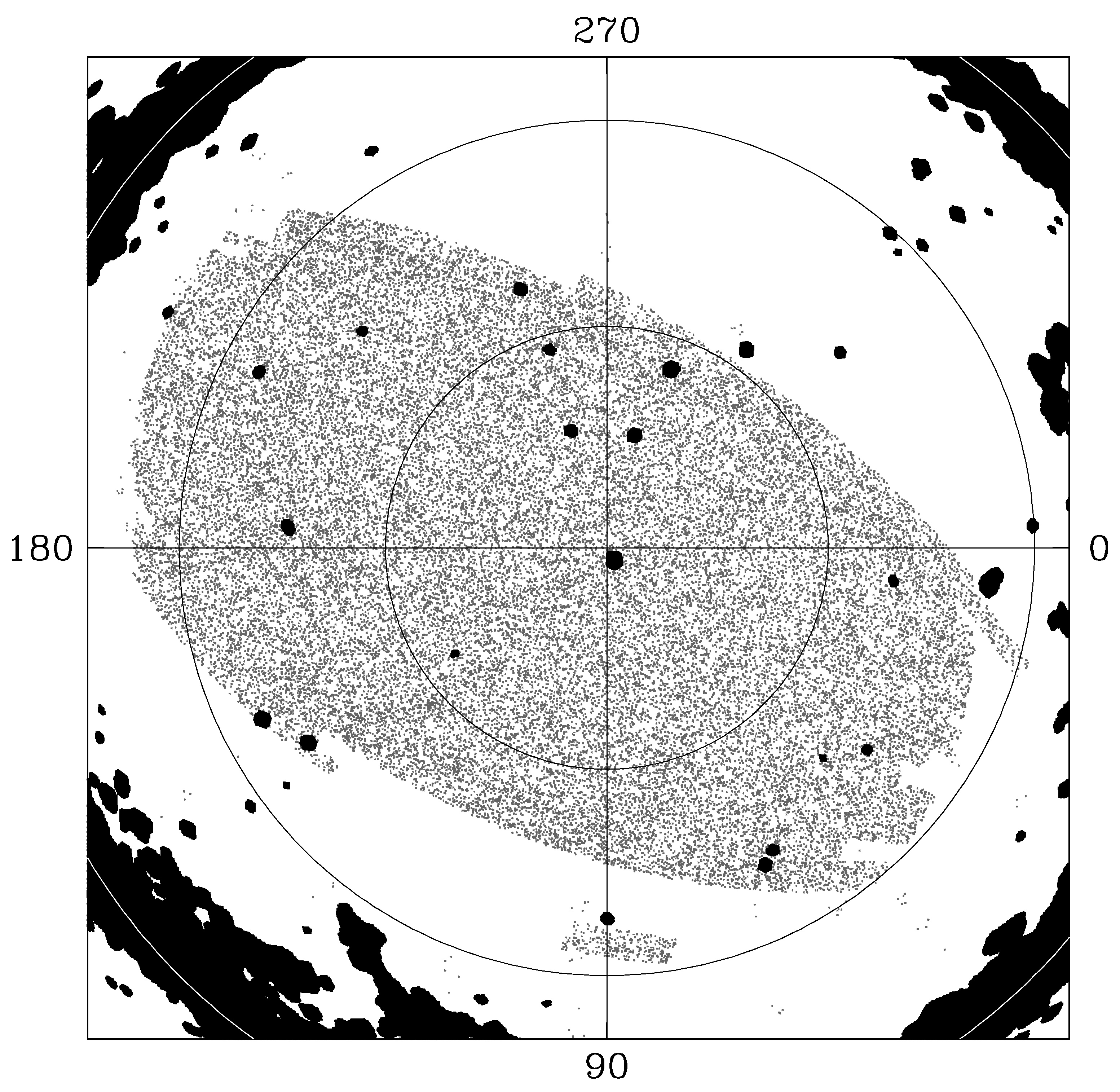}
   \caption{Distribution of $\sim 42400$ quasars  brighter than $z = 19.50$
     in the north galactic hemisphere from the SDSS DR7 quasar catalogue
     at distances between $1300$ and $4700$\,Mpc. {\it Planck's} SEVEM mask
     is superimposed in black (see Sec.~\ref{sec:cmb_map}).
     Galactic latitude circles of $60^\circ$, $30^\circ$ and $0^\circ$
     (arcs) are marked.}
   \label{fig:qso_map}
\end{figure}

\begin{figure}
   \includegraphics[width=1.00\linewidth]{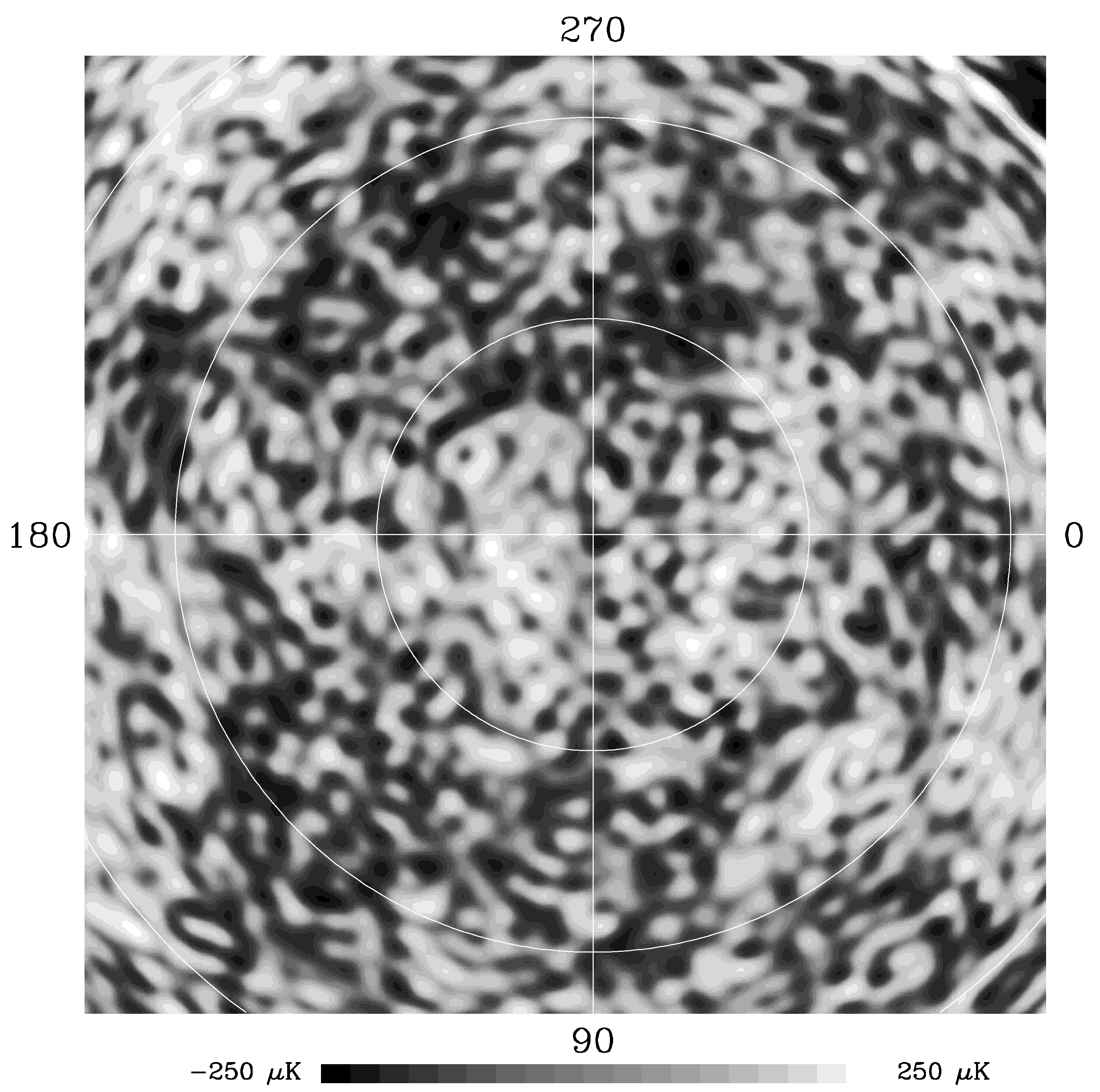}
   \caption{Spherical harmonic expansion of the
   {\it Planck} $2015$ SEVEM map in the north galactic hemisphere
   in polar projection (same area as Fig.~\ref{fig:qso_map}) up
   to order $l = 90$.}
   \label{fig:cmb_map}
\end{figure}

The Sloan Digital Sky Survey Quasar Catalogue covers a huge volume of space
that allows for statistical investigation of the relationship between the large
scale matter distribution and the fluctuations of the Cosmic Microwave
Background (CMB) induced by the ISW effect. The fifth
edition of this catalogue is described in detail by \citet{schneider10}.  All
the procedures of the  quasar selection, photometry and spectroscopic redshifts
are presented in that paper. Here we summon only the overall parameters of the
catalogue.  We used the data in the north galactic hemisphere (NGH). Total number of
quasars in this area exceeds $90\,000$.  Additionally, we impose a rigid
magnitude limit  of $19.5$ in the $z$ band. Albeit, this $z$ cut-off decreases
the number of quasars in the northern hemisphere to $\sim 70\,000$, it
effectively reduces variations of the quasar surface density in the selected
deep areas.  Despite great effort to achieve statistical homogeneity of the
catalogue, objects satisfying the selection criteria are subject to some
residual biases discussed in detail by \citet{schneider10} (see also
\citealt{soltan17}). In the present investigation we concentrate on the
correlation between the quasar space distribution and the fluctuations of the
CMB temperature. Admitting that various features of the catalogue generated
locally in the process of the data acquisition increase the uncertainties of
the final results, one can expect that these biases do not affect amplitudes of
the cosmic signal. This conclusion applies also to our distance estimates. We
ignore deviations from the Hubble flow and the distance is derived from the
Hubble relationship in the considered cosmological model.

Distributions of quasars at distances between $1300$ and $5200$\, Mpc ($0.33
\lesssim z \lesssim 2.01$) is shown with the histogram in
Fig.~\ref{fig:distance_fit}. The curve represents a fit by Legendre polynomial
of degree $n=50$. The fit is used to generate quasi-random distribution of
points using the MC method (see below).  The catalogue is magnitude limited
what introduces an overall gradient of object space density.  Although, a
strong cosmic evolution of quasars reduces to some extent the amplitude of this
effect, the average space density of the catalogued objects at distance of
$4500$\,Mpc is $\sim 4$ times lower than at $1500$\,Mpc.  A wide plateau
between $3000$ and $4500$\,Mpc comes from  a kind of interplay between the
observational selection and the evolution. The distribution of the catalogued
objects in the celestial sphere in galactic coordinates is shown in
Fig.~\ref{fig:qso_map}. A polar projection is used. Despite its
featureless appearance, the subsequent analysis shows that the space
distribution of objects is statistically nonuniform, and both concentrations of
quasars and the under-dense regions reflect the large scale fluctuations
of matter that in turn generate the ISW signal.

\subsection{CMB map \label{sec:cmb_map}}

The CMB temperatures data are based on the {\it Planck} 2015 data release
\footnote{http://pla.esac.esa.int/pla/\#maps} \citep{planckXI}.  Four maps:
Commander, NILC, SEVEM and SMICA, that differ in methods of background
subtraction are available. We use the SEVEM maps. One of the objectives of the
SEVEM method was to minimize the variance of the clean map outside the
confidence mask. This was achieved in a two step procedure. First, a set of
template maps with removed the CMB signal was constructed. Then, a linear
combination of templates was subtracted from the CMB-dominated maps.  Since the
expected amplitude of the ISW signal is substantially smaller than the integral
fluctuations of the registered flux, the condition of minimum variance seems to
be essential in our investigation. One should note also that in several
investigations of the ISW effect, e.g.  \citet{planckXIX,planckXXI,nadathur16}
the results using all the maps are very much alike.  We used standard Res 10
HEALPix\footnote{https://healpix.sourceforge.io/} projection \citep{gorski05}
with $\sim 3.4$ arcmin resolution. The NGH section of SEVEM
data smoothed with the spherical harmonic filter of degree $l = 90$ is shown in
Fig.~\ref{fig:cmb_map}.  In all calculations the confidence mask leaving
approximately $85$ per cent of useful data has been applied. It is shown in
Fig.~\ref{fig:qso_map}.


\section{Quasars and the ISW}
\label{sec:qso_isw}

\subsection{The ISW effect in $\Lambda$CDM \label{sec:isw}}

Here we collect the formulae that describe effects of gravitational potential
variations on the CMB photons in the low redshift Universe. We limit the
analysis to the linear large-scale fluctuations of matter distribution in the
$\Lambda$CDM model, known as the late-time Sachs-Wolfe effect. The temperature
$T({\bf \hat p})$ of the CMB propagating in the direction defined by a unit
vector $\bf \hat p$ deviates from the average temperature $\overline T$
\citep[p.\,238\footnote{The last term in Equation 8.56.}]{dodelson03}:

\begin{equation}
\frac{\dT({\bf \hat p})}{\overline T}\, =\, 
   \frac{T({\bf \hat p})- \overline T}{\overline T}\, =\, \frac{2}{c^2}\,
   \int_{\eta_{\rm LS}}^{\eta_0}
           \rmd\eta\; \frac{\rmd \Phi({\bf r},\eta)}{\rmd\eta}\,,
 \label{eq:isw_1}
\end{equation}

\ni where $\eta$ is the conformal time, ${\rm LS}$ and $0$ denote the last
scattering surface and the present moment, respectively; $\Phi({\bf r},\eta)$
is the distribution of the Newtonian gravitational potential along the photon
path, ${\bf r}\, =\, {\bf \hat p}\, r$, where $r$ is comoving distance, and $c$
is speed of light.

Below we will model the gravitational potential distribution using the quasar
sample, and it is convenient to rewrite Eq.~\ref{eq:isw_1} in the form:

\begin{equation}
\frac{\dT({\bf \hat p})}{\overline T}\, =\, \frac{2}{c^3}\,
    \int_{0}^{r_{\rm LS}} \rmd r\, a\, \Dot{\Phi}({\bf r},\,t)\,,
 \label{eq:isw_2}
\end{equation}

\ni where $\bf{\hat p}$ is now unit vector defining direction in the sphere,
$a = a(t)$ -- scale factor of the universal expansion, and
overdot denotes a time derivative.

In the matter-dominated flat Universe, i.e. with the critical density generated
exclusively by matter, $\Omega_{\rm m}\, =\, 1$, linear density fluctuations
develop at the same rate as the scale factor $a$. Thus, the proper linear size
of individual structure is proportional to its mass and gravitational potential
does not evolve in time.  Higher rate of the  Universe expansion due to non-zero
cosmological constant removes this degeneracy, and induces time evolution of
$\Phi$.

A standard procedure to bind matter density fluctuations with the gravitational
potential is to use the Poisson equation in Fourier space (e.g.
\citealt{nadathur12}):

\begin{equation}
\Phi({\bf k}, t)\, =\, -\frac{3}{2}\,H_0^2\, \Omega_{\rm m}\,
   \frac{\delta({\bf k}, t)}{k^2\, a}\,,
\label{eq:phi}
\end{equation}

\ni where $\delta({\bf k}, t)$ is the Fourier transform of the matter density
fluctuations:

\begin{equation}
\delta({\bf r}, t)\, =\, \frac{\rho({\bf r}) - \overline \rho}{\overline \rho}\,.
\label{eq:delta}
\end{equation}

The evolution of density fluctuations in the linear regime describes a
growth factor $D(t)$: $\delta({\bf k}, t)\, =\, D(t)\, \delta_0({\bf k})$.
Time derivative of the Potential $\Phi$ is defined entirely by this function:

\begin{equation}
\dot \Phi({\bf k}, t)\, =\, -\frac{3}{2}\,H_0^2\, \Omega_{\rm m}\,
   \frac{\delta_0({\bf k})}{k^2}\, \left[ a\,\frac{\rmd D}{\rmd a} - D \right]\,
   \frac{\dot a}{a^2}\,,
\end{equation}

\ni and substituting $\beta(t)\, =\, \rmd \ln D / \rmd \ln a$ we get
(\citealt{nadathur12}):

\begin{equation}
\dot \Phi({\bf k}, t)\, =\, \frac{3}{2}\,H_0^2\, \Omega_{\rm m}\,
   \frac{H(t)}{a}\, \frac{\delta({\rm k}, t)}{k^2}\,
   \left[1 - \beta(t)\right]\,.
\label{eq:phi_k}
\end{equation}

\ni In the subsequent calculations, the relevant terms are derived for the
$\Lambda$CDM cosmological model with parameters specified in the Introduction.
The formulae for the linear growth rate, $D(t)$ were taken from
\citet[][p.\,49--51]{peebles80}.

If along the photon path there is a single spherically symmetric structure,
the inverse transform of $\dot \Phi({\bf k}, t)$ is given in a closed form:

\begin{equation}
\dot \Phi({\bf r},t)\, =\, \frac{3}{2}\,H_0^2\, \Omega_{\rm m}\,
     \frac{H(t)}{a}\, \left[1 - \beta(t)\right]\, F(s)\,,
\label{eq:phi_dot}
\end{equation}

\ni where $s$ is now distance form the structure centre, and 

\begin{equation}
F(s)\, =\, \frac{1}{s}\, \int_0^s \rmd s^\prime\, {s^\prime}^2\,
 \delta(s^\prime, t)\, +\,
    \int_s^\infty \rmd s^\prime\, s^\prime\, \delta(s^\prime, t)\,.
\label{eq:f_s}
\end{equation}

In the linear case, the same formalism may be applied to the set of (spherical)
objects distributed along the line of sight. In this case, the density profile
$F(s)$ is replaced by a sum of profiles that represent separate structures.
Albeit complexity of the cosmic matter distribution is not adequately
represented by such simple arrangements, we are going to adapt the above
formula in order to seek the correlation between the quasar distribution and
the CMB map. In the following section we construct a simple statistics
using the spatial distribution of quasars. Although, it cannot serve as the
estimator of the gravitational potential, it is correlated with the potential.
Then, this statistics is substituted into Eqs~\ref{eq:phi_dot} and
\ref{eq:f_s}, and integrated according to Eq.~\ref{eq:isw_2}.

\subsection{Quasar statistics vs. CMB temperature fluctuations
\label{sec:qso_phi}}

A problematic application of the quasar sample to the investigation of the ISW
effect depends on the relationship between the two highly disparate
distributions: discrete, scarce quasar population and continuous gravitational
potential. If a mean separation between the tracer objects is comparable
or smaller than the interesting linear scale of potential variations, then the
matter density field may be efficiently reconstructed using the Voronoi
tessellation technique applied to the discrete data (e.g.
\citealt{vandeweygaert09}). This method was effectively utilized by by
\citet{granett09}, who modelled the potential using a sample of $400\,000$
galaxies at $z \sim 0.5$ from the SDSS (\citealt{adelman08}).

\citet{nadathur12} showed that within the linear approximation in the
$\Lambda$CDM model, the ISW signal generated by numerous small and moderate
size structures is much weaker than the primordial fluctuations of the CMB
occurring at the surface of last scattering.  It appears that only the
superstructures of size $\sim\!100$\,Mpc or larger produce signal that is
statistically distinguishable from the background.  The mean distance between
the neighbouring galaxies in the \citet{granett09} investigation amounts to
$\sim\!19$\,Mpc. Thus, spatial resolution attainable by means of the Voronoi
tessellation is adequate to estimate potential on the required scales.  The
mean distance in our quasar sample approaches $100$\,Mpc at $z\,=\,0.5$ and
increases to $124$\,Mpc at $z\,=\,1$. Because of the much lower space
density, the discrete nature of the distribution strongly
affects the local density estimates. We assess that a less refined approach
than the Voronoi tessellation would be adequate.

One can expect that Poisson noise in our sample will hamper an identification
of individual superstructures in the quasar catalogue.  Nevertheless, a
relationship between large scale matter distribution and quasars exists still
in the form of statistical correlation of the local number density of quasars
with the total matter density. It implies that in the randomly selected volume
of space, $V$, number of quasars is correlated with the total amount of matter
contained in that volume. To keep a linear character of the correlation we make
a standard assumption that a `quasar bias factor', i.e. the ratio of the
relative overdensities of quasars to total matter is constant in statistical
sense. The amplitude of this correlation, or correlation coefficient, depends
critically on space density of quasars, and it is expected to be strongly
reduced by the stochastic nature of the Poisson distribution. Keeping in mind this
limitation, we define the amplitude of quasar space density variations the same
way as the total matter density fluctuations:

\begin{equation}
\delta \nu\, =\, \frac{n_{\rm o} - n_{\rm f}}{n_{\rm f}}\,,
\label{eq:deltan}
\end{equation}

\ni where $n_{\rm o}$ and $n_{\rm f}$ are the observed and average number of
quasars in $V$. Because of a strong radial density gradient in our quasar
sample, excess or deficiency of quasars is normalized to the average number of
quasars expected for given distance, $n_{\rm f} = n_{\rm f}(D)$. The average
number is determined using the Lagrange fit shown in
Fig.~\ref{fig:distance_fit}. In all the following considerations, the test
volume $V = V(r_{\rm s})$ is a sphere, and the interesting correlations will be
investigated over a wide range of the sphere radii, $r_{\rm s}$.

\begin{figure}
   \includegraphics[width=1.00\linewidth]{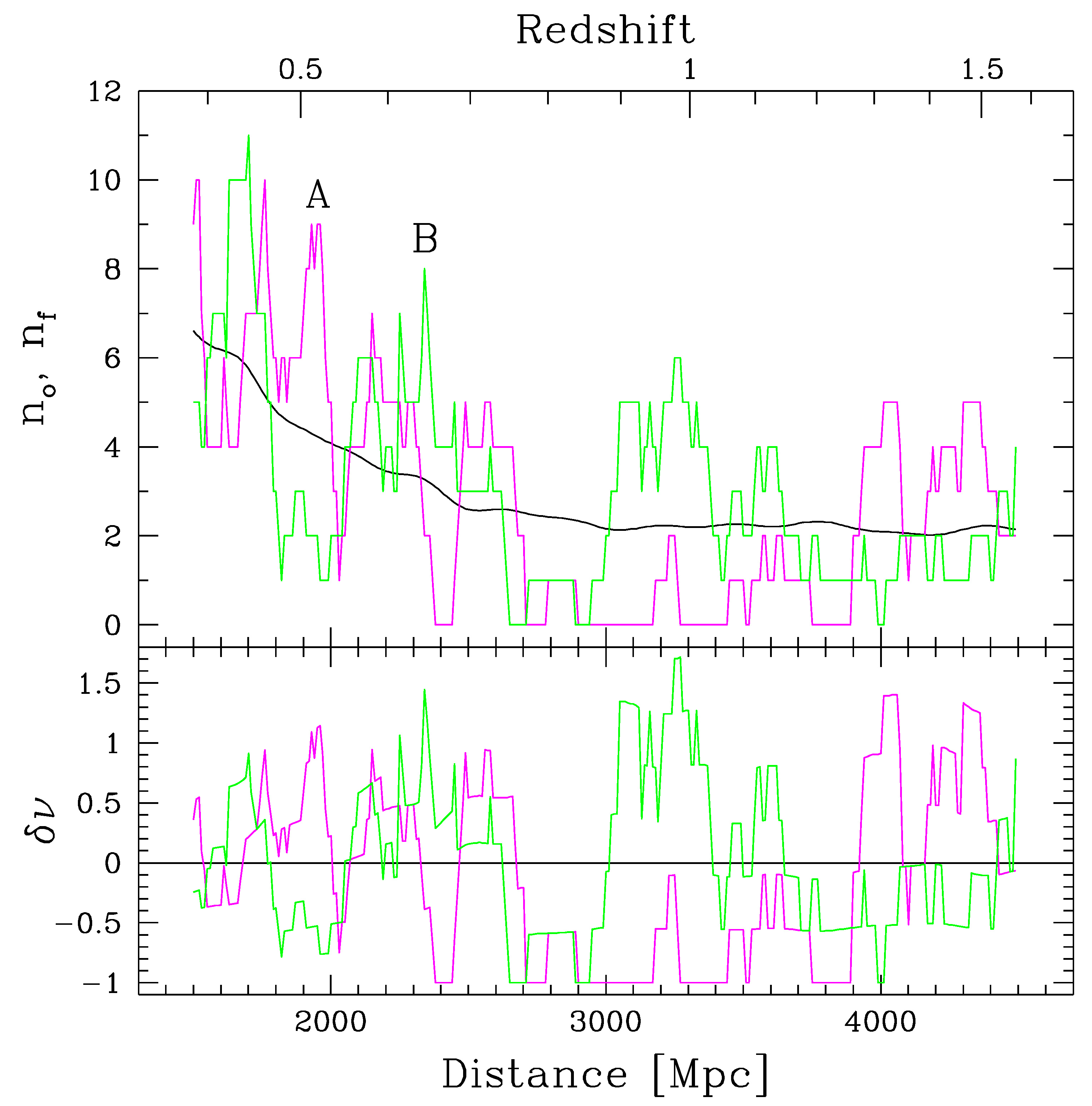}
   \caption{Two typical radial distributions of quasars along radial
   directions: A($l=58^\circ,b=50^\circ$) and B($l=236^\circ,b=60^\circ$),
   for sphere radius $r = 100$\,Mpc (see text). Upper panel: observed
    numbers of quasars, $n_{\rm o}$ (erratic curves), and the average number 
   based on the Legendre fit, $n_{\rm f}$. Lower
   panel: relative fluctuations of quasar concentration as defined in
   Eq.~\ref{eq:deltan}.} 
   \label{fig:radial_trends}
\end{figure}

In Fig~\ref{fig:radial_trends} we show two typical distributions of quasars
between $1500$\,Mpc and $4500$\, Mpc.  At each point objects are counted within
a sphere of $100$\,Mpc radius. Plots exemplify typical patterns observed in the
catalogue.  Upper panel gives the actual numbers of observed objects, what
illustrates importance of the Poisson scatter.
   
Cosmological simulations allow to relate distribution of tracing objects, with
the total matter density, then - with the structure of the gravitational
potential, and the ISW effect (e.g.  \citealt{watson14a,nadathur17}).
Observational data do not provide such gratifying information. Therefore we
seek for a direct {\it empiric} relationship between the quasar distribution
and the ISW signal. It is evident  that the $\delta \nu$ parameter
constitutes a poor indicator of the potential. We notice, however, that under
the assumption of the constant quasar bias factor, the gravitational potential
$\Phi({\bf r},t)$ along the photon path is correlated with the distribution of
$\delta \nu = \delta \nu({\bf r})$.  In the linear approximation the potential
is also linearly correlated also with the local matter density fluctuation
averaged over a sphere of finite radius $r_{\rm s}$.  Since the CMB temperature
variations are related to $\Phi({\bf r},t)$ via Eq.~\ref{eq:isw_2}, one can
expect also the correlation between $\delta \nu$ and $\dT$. One should
emphasize that no definite relationship between $\delta \nu$ and $\Phi({\bf
r},t)$ or $\dT$ is assumed. In particular, the quasar bias factor is not
present in the calculations.

As we are interested in the impact of the variations of the matter distribution
on the CMB, one should consider the matter density fluctuations over the whole
section of the photon paths within the selected volumes. Accordingly, we define
the `net' density deviation between distances $D_{\rm l}$ and $D_{\rm h}$ as:

\begin{equation}
\DN\, =\, \frac{1}{D_{\rm h} - D_{\rm l}}\,
         \int_{D_{\rm l}}^{D_{\rm h}} \delta \nu\, \rmd D\,.
\label{eq:Delta_q}
\end{equation}

One can expect that the distribution of the $\DN$ over the sky should
correlate with the ISW signal. Thus, $\DN$ should also correlate with with
the total CMB variations. Yet, the functional relationships between the matter
density distributions and the gravitational potential given in
Eqs.~\ref{eq:phi_dot} and \ref{eq:f_s} indicate that the stronger correlation
is expected when the local density fluctuation $\delta\nu$
is substituted into Eq.~\ref{eq:phi_dot} instead of $F(s)$.
The modified form of Eq.~\ref{eq:phi_dot} is substituted into
Eq.~\ref{eq:isw_2}, that defines a new parameter $\DQ$ is as follows:

\begin{equation}
\DQ\, =\, 3\,\frac{\overline T}{c^3}\, H_0^2\, \Omega_{\rm m}
   \int_{D_l}^{\rm D_h} \rmd r\, 
     H(t) \left[1 - \beta(t)\right]\, \delta\nu\,.
\label{eq:Delta_Q}
\end{equation}

\ni Both $\DN$ and $\DQ$ are defined as the averaged density fluctuations
$\delta \nu$. The $\DN$ is a simple arithmetic average over distance, while in
$\DQ$ the fluctuations are weighted in the same way as the gravitational
potential in the ISW effect. Although amplitude $\DQ$ is measured in $\mu$K,
this quantity should not be treated as the estimate of the ISW effect generated
by the matter distribution between $D_{\rm l}$ and $D_{\rm h}$. $\DQ$ is a
variable functionally dependent on the quasar distribution, and as such it is
expected to correlate with the the ISW signal. 

\begin{figure}
   \includegraphics[width=1.00\linewidth]{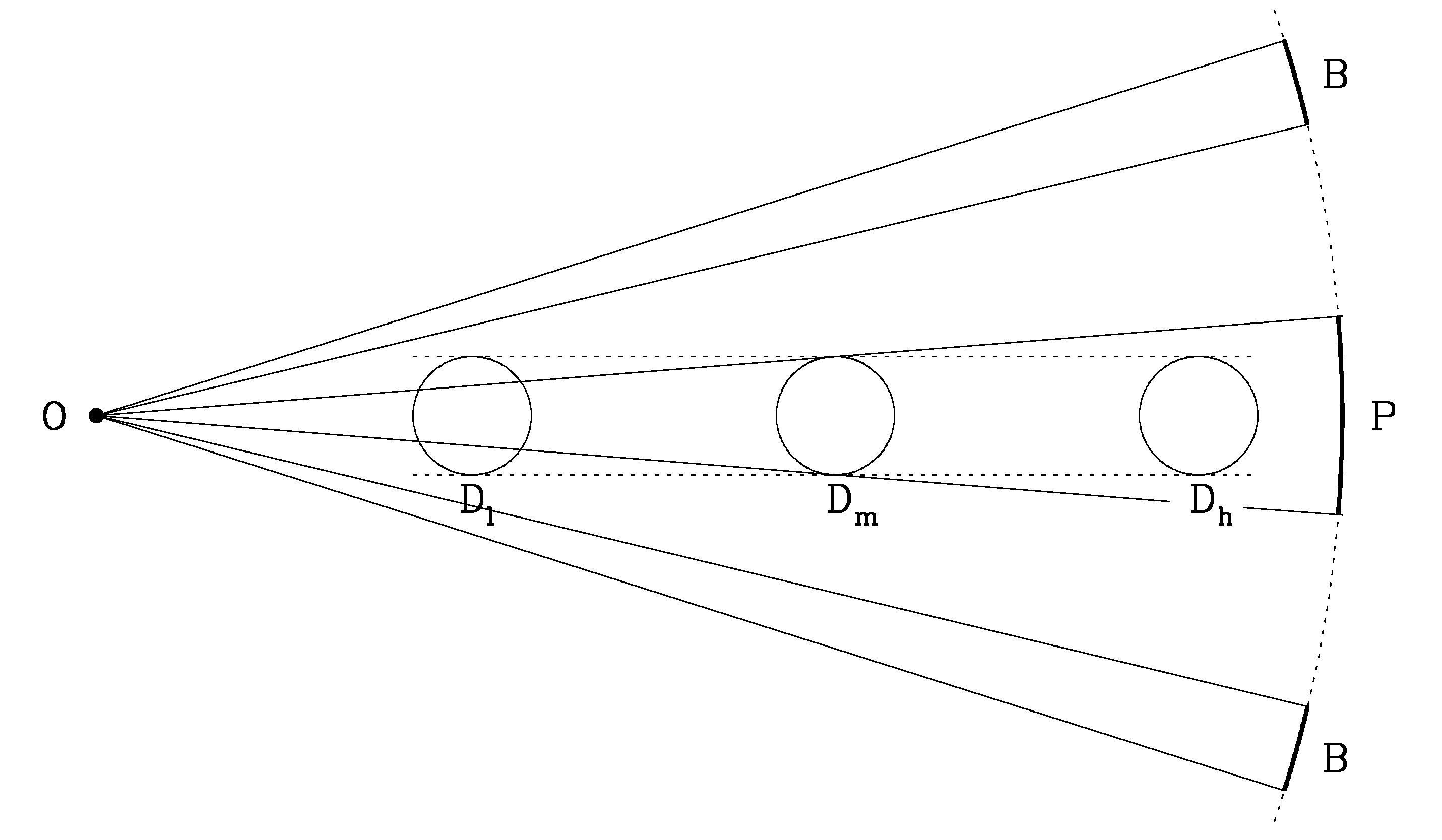}
   \caption{Schematic view indicating configuration of the areas
   used to determine amplitudes $\DN$ i $\dT$ for a single pointing.
   Quasars within a 3D volume swept through by the sphere moving
   between $D_{\rm l}$  and $D_{\rm h}$ contribute to the $\DN$ estimate,
   while 2D areas `P' and `B' are used to determine $\dT$ fluctuation
   (see text).}
   \label{fig:settings}
\end{figure}

We analyze three subsets of the quasar catalogue: the whole data between
$D_{\rm l}=1500$\,Mpc and $D_{\rm h}=4500$\,Mpc, and separately quasars in two
distance bins: $1500-3000$\,Mpc and $3000-4500$\,Mpc.  To assess the
correlation coefficient between $\DQ$ and $\dT$, we consider a dense net of
equally spaced pointings over the area covered by the SDSS quasar catalogue.
Spacing between pointings was selected at $\theta = 60$\,arcmin (see below).
For each pointing the amplitudes of $\DQ$ and the CMB temperature signal are
calculated. Geometric settings for the data acquisition are shown in
Fig.~\ref{fig:settings}. The quasar relative density variations, $\DQ$, are
determined between the distances $D_{\rm l}$ and $D_{\rm h}$. A question of the
angular extent of the individual $\dT$ measurement is not well-defined.  First,
the definition of $\DQ$ variable neither determines the characteristic
angular scale for the gravitational potential variations, nor defines the
optimal shape of the filter. Second, $\DQ$ is determined using fixed volume
$V=V(r_{\rm s})$ at a wide range of distances.

With no direct constraints on the optimal size for the temperature evaluation
area, we take a 2D Gaussian function parametrized by variance $\sigma_{\rm
P}^2$.  It is reasonable to assume that the angular scale of $\dT$ estimates
should correspond to the `average' angular size of the area used to calculate
$\DQ$.  This is because the area of the highest amplitude of the ISW signal
roughly coincides with the structure size. A radius of this 'reference' area
(denoted $P$ in Fig.~\ref{fig:settings}) $\vartheta = r_{\rm s} / D_{\rm m}$,
where $D_{\rm m}$ is the median distance between $D_{\rm l}$ and $D_{\rm h}$.
Consequently, the temperature assigned to the pointing is the average
temperature weighted with the Gaussian function of width $\sigma_{\rm P}$.
Geometrical settings are insufficient to specify $\sigma_{\rm P}$ and it is
considered a free parameter. In the calculations several values of $\sigma_{\rm
P}$ were adopted within a range

\begin{equation}
0.2\,\vartheta\, \leq\, \sigma_{\rm P}\, \leq \, 1\,\vartheta\,,
\label{eq:filter}
\end{equation}

\ni The radius of the integration area was assumed at $3\,\sigma_{\rm P}$
for each $\sigma_{\rm P}$.

To minimize the large angular scale temperature variations uncorrelated with
$\DQ$, $\dT$ is defined as a difference between the temperature in the pointing
direction and in the surrounding area (denoted by B in
Fig.~\ref{fig:settings}), defined as a ring centered on given pointing with the
inner radius $\theta_{\rm in}\, =\, \max[1.5\,  r_{\rm s} / D_{\rm l},\,
3\,\sigma_{\rm P}]$\,, and width of $4$\,deg. Thus, procedure to determine
$\dT$ corresponds to the use of Mexican hat filter with flat negative section.
All the calculations were performed for sphere radii between $40$ and
$140$\,Mpc.


\section{Contribution of the ISW effect to the CMB maps}
\label{sec:results}

\begin{figure}
   \includegraphics[width=1.00\linewidth]{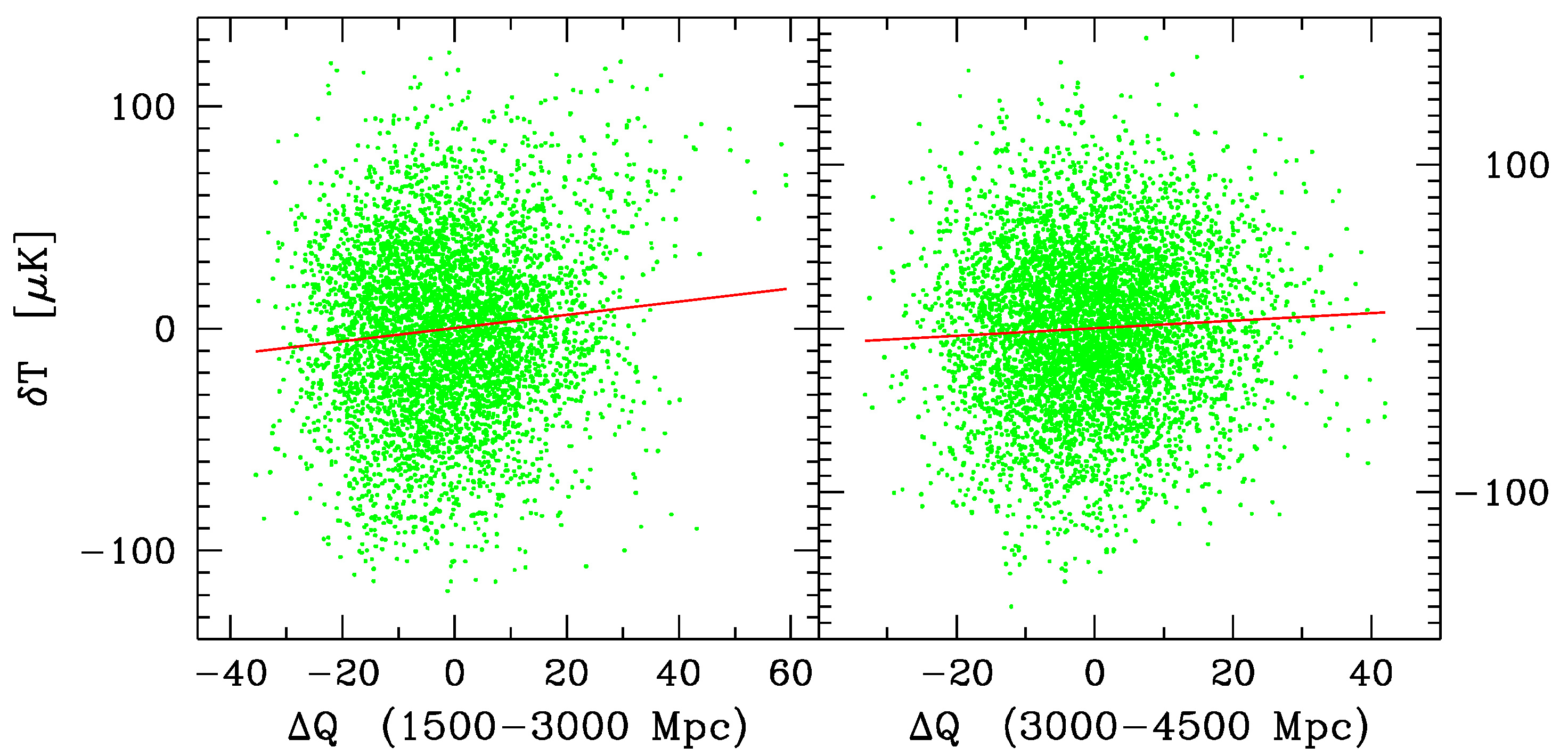}
   \caption{Distributions of local temperature amplitude and the quasar excess
   density in the distance bin $1500 - 3000$\,Mpc (left panel), and
   $3000 - 4500$\,Mpc (right panel)  for spheres of radius
   $r_{\rm s} = 100$\,Mpc, and $\sigma_{\rm P} = 0.4\,\vartheta$
   in more than $5000$ pointings  (see text). Tilted lines are the least
   square best fits of $\dT$ on $\DQ$.}
   \label{fig:phi_dt}
\end{figure}

High concentrations of pointings caused by their small separations ensured
efficient exploitation of all the features of the QSO catalogue. Total number
of useful pointings, $N_{\rm pnt}$, i.e. pointings not affected by the mask,
and sufficiently distant from the survey edges to warrant correct $\delta \nu$
counts, depends on the minimum distance, $D_{\rm l}$, and the sphere radius,
$r_{\rm s}$. For $D_{\rm l} = 1500$\,Mpc and $r_{\rm s} = 40$\,Mpc it exceeds
$5600$, and drops to $\sim 4500$ for $r_{\rm s} = 140$\,Mpc. In
Fig.~\ref{fig:phi_dt} we plot  the local temperature fluctuation, $\dT$, for
$\sigma_{\rm P} = 0.4\,\vartheta$, against the relative density excess, $\DQ$,
in more than $5000$ pointings for two distance bins $D_{\rm l} - D_{\rm h}$:
$1500 - 3000$\,Mpc and $3000 - 4500$\,Mpc, for $r_{\rm s} = 100$\,Mpc.  The
data are positively correlated. The Pearson correlation coefficients $\rho =
0.0992$ and $0,0528$ for the low and high distance bin, respectively. For the
merged data, i.e. between $1500$ and $4500$\,Mpc the correlation coefficients
amounts to $0.1082$. We calculated also correlation coefficients for the
$\dT - \DN$ distribution.  In agreement with the reasoning above, the figures
in this case are systematically lower, but the differences are insignificant.
For the low, high and merged distance bins the corresponding coefficients are
$0.0933$, $0.0522$ and $0.1038$, respectively.

Although the cloud of points is adequately represented by a 2D Gaussian
distribution, a significance of the correlation is not given by standard
statistical formulae applicable for the Gaussian function. Because of tight
pointing spacing, the amplitudes $\DQ$ are not independent variables, and the
uncertainty range of $\rho$ has to be assessed separately using the MC
simulations.

\begin{figure}
   \includegraphics[width=1.00\linewidth]{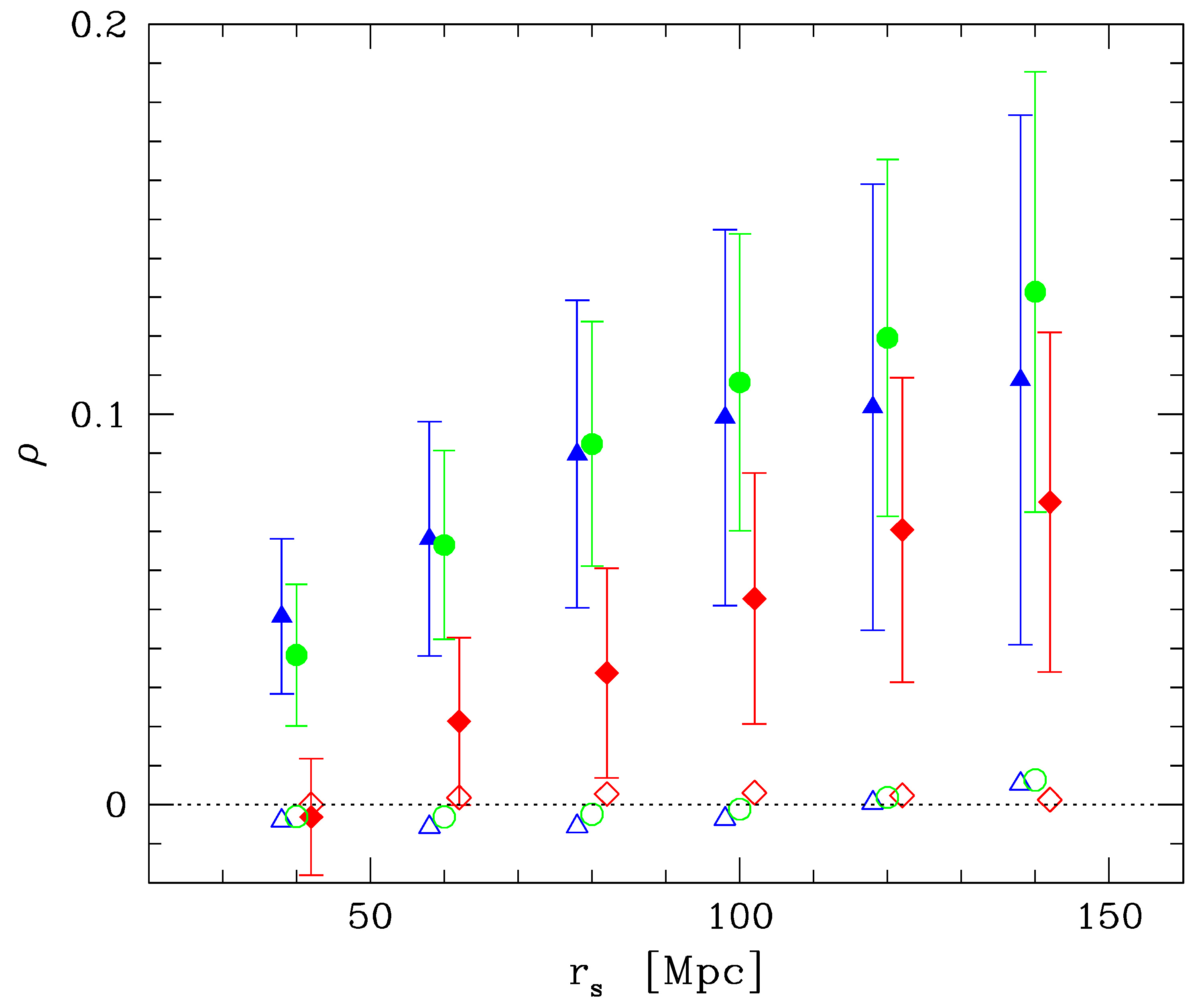}
   \caption{The Pearson correlation coefficient, $\rho_{\rm N}$, between $\dT$
   for $\sigma_{\rm P} = 0.4\,\vartheta$,
   and $\DN$ in the distance bin $1500 - 3000$\,Mpc (full triangles),
   $3000 - 4500$\,Mpc (full squares), and $1500 - 4500$\,Mpc (full circles)
   for a range of sphere radii
   $r_{\rm s}$; error bars represent rms scatter of the correlation
   coefficients computed for $55$ sets of mock data (see text). Open symbols
   - average coefficients computed for the mock data.}
   \label{fig:rho_phi}
\end{figure}

We calculate correlation coefficients for a large number of the mock data. The
SDSS QSO and CMB temperature maps are shifted against each other to erase the
physical correlation.  The rms scatter of the  residual correlation signal
calculated for combined map pairs is taken as representing the statistical
noise of our estimate of $\rho$. To get sufficiently large number of combined
map pairs the following procedure is applied. First, we rotate the quasar data
in $15^\circ$ steps about  the axis $(l, b) = (120^\circ, 0^\circ)$. In
galactic coordinate system it shifts quasars at the north galactic pole in
Fig.~\ref{fig:qso_map} south along longitude of $30^\circ$.  This method
generates $23$ sets of uncorrelated quasar and CMB maps.  Then, the quasar data
are rotated by $60^\circ$ about the galactic polar axis and the procedure to
rotate the map in  $15^\circ$ steps, this time about $(l,b) = (120^\circ,
0^\circ)$ axis, is repeated. Finally, the quasar data are rotated by
$120^\circ$ about the galactic polar axis and $15^\circ$ step map
multiplication about $(l,b) = (180^\circ, 0^\circ)$ axis is executed. This
scheme generates in total $69$ data sets. For most of data pairs some pointings
fall in the masked area in the SEVEM map. If more than half of the area
used to estimated $\dT$ is masked out, the pointing is not used. Therefore the
correlation coefficients of the test data are determined using the smaller
number of pointings. For this reason, estimates of the $\rho$ rms get
larger above the level expected for the real data. To lessen this effect, the
rms was calculated using only the test maps with the number
of accepted pointings larger than $75$ per cent of pointings used in the real
data. Removal of heavily masked maps had minimal effect on error estimates.
Within the considered range of all the parameters the number of the
test data sets used in the error estimates has not dropped below $54$.

In Fig.~\ref{fig:rho_phi} we plot the $\dT - \DQ$ correlation coefficients in
all three distance bins for a range of sphere radii.  The error bars are
estimated using the mock data sets.  For all the combinations of parameters,
the correlations are positive. However the signal in the $3000 - 4500$\,Mpc bin
is substantially weaker and never exceeds  $2\,\sigma$.  In the full distance
range of $1500 - 4500$\,Mpc  the positive $\dT - \DQ$ correlation is
significant above  $2.5\,\sigma$ for the sphere radii $50 \le r_{\rm s} \le
120$\,Mpc.  The data divided into low and high distance bins show weaker
correlation significance  but in the $1500 - 3000$\,Mpc bin within the range
$40 \le r_{\rm s} \le 100$\,Mpc it still exceeds $2\,\sigma$.

Moderate, but measurable correlations between $\dT$ and $\DQ$ provide
intriguing constraints on the amplitude of the ISW effect produced by the
matter accumulated between $D_{\rm l}$ and $D_{\rm h}$.  For the purpose of the
present analysis the total observed temperature deviation from the average,
$\dT$, is a sum of four components: primordial fluctuations generated in the
recombination era, the ISW effect produced in the $D_{\rm l} -D_{\rm h}$
distance range, the ISW outside these distances, and measurement errors.
Clearly, only the ISW effect generated between  $D_{\rm l} -D_{\rm h}$ can
correlate with $\DQ$.  Thus, the temperature fluctuations $\dT$ may be
decomposed into two statistically independent parts: $\dT\, =\, \dT_{\rm un} +
\dT_{\rm cr}$, uncorrelated and correlated with the $\DQ$, respectively.

Let $t$ and $q$ denote the amplitudes of $\dT$ and $\DQ$ for individual
pointings. In Appendix~\ref{app:corr}  we recall canonical relationships
between standard deviations of the linearly correlated variables, and apply the
respective formulae to the quantities  $\sgt$ and $\sgq$. We find also the
relationship between these quantities and the standard deviation of the ISW
effect. The relationships are obtained under the assumption that the variables
$t$ and $q$ have the averages $\mu_{\rm t}$ and $\mu_{\rm q}$ equal to zero and
the Pearson correlation coefficient $\rho_{\rm tq} \neq 0$.

The ISW effect induced by the matter density fluctuations between $D_{\rm l}$
and $D_{\rm h}$ generates the rms temperature fluctuations, $\sgs$ that
contribute to the total CMB temperature variations $\sgt$. According to
Eq.~\ref{eq:sigmas}, the variance $\sgssq$  is a sum of two positive
components: $\sigma_{s|c}^2$ and $\sgosq$, where the former term,
$\sigma_{s|c}$, represents the scatter that is not directly observable (see
Appendix~\ref{app:corr}), while the latter one can be determined from the data.
Therefore, the present method provides the lower limit for the ISW effect
(Eq.~\ref{eq:sig_o}):

\begin{equation}
\sgs\, >\, \sgo\, =\,
 \rho_{\rm tq}\,\frac{\sgt\,\sgq}{\sqrt{\sgqsq\,-\,\sgrsq}}\,,
\label{eq:isw_ll}
\end{equation}

\ni where $\sgr$ represents a contribution to the $\sgq$ generated by
stochastic nature of the discrete quasar data. The amplitude of variable $\sgr$
is determined by the Poisson statistics. Consequently, $\sgr$ is correlated
neither with the ISW effect nor the total CMB fluctuations. To assess $\sgr$, a
large number of the mock quasar data were generated using the Monte Carlo
scheme in which the spherical coordinates were randomized while the radial
coordinate of points were drawn from the probability distribution function
based on the fit shown in Fig.~\ref{fig:distance_fit}.
Equation~\ref{eq:isw_ll} provides limits for the ISW effect based on
correlations in sky coordinates. Its counterpart in the harmonic space are
given by \citet{granett09}.  The lower bound rather than the actual amplitude
of the ISW signal results from the fact that no definite relationship between
the gravitational potential and the quasar distribution has been used.

\begin{figure}
   \includegraphics[width=1.00\linewidth]{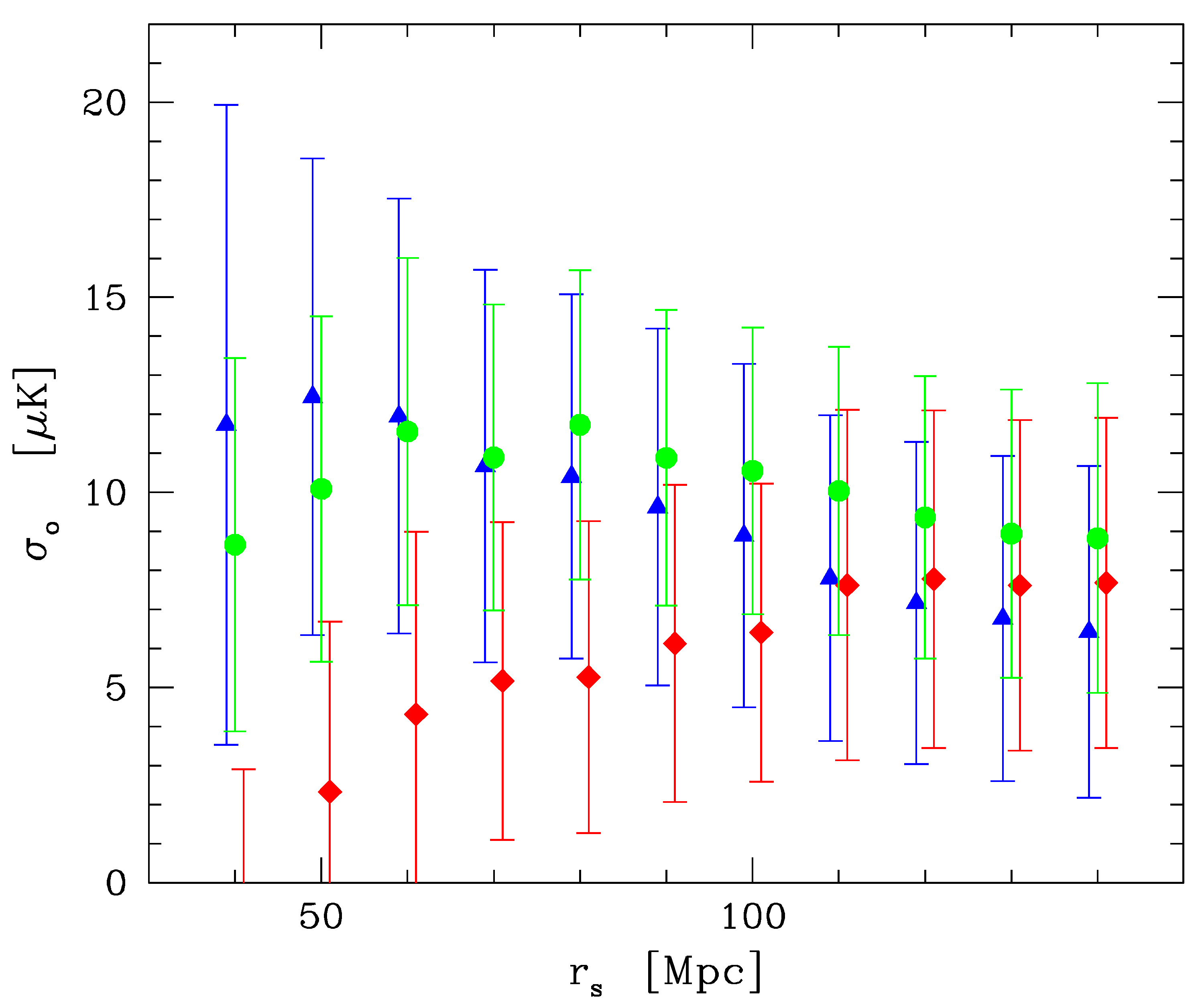}
   \caption{Observational lower limits for the rms fluctuations of the
   ISW signal (Eq.~\ref{eq:sig_o}) for distances $D_{\rm l} - D_{\rm h} =
   1500 - 3000$\,Mpc (triangles), $3000 - 4500$\,Mpc (diamonds),
   and $1500 - 4500$\,Mpc (dots), for sphere radii
   $40 \leq r_{\rm s} \leq 140$\,Mpc, and the filter parameter
   $\sigma_{\rm P}\,=\,0.4\,\vartheta$ (Eq.~\ref{eq:filter}).}
   \label{fig:rms_o_rs}
\end{figure}

\begin{figure}
   \includegraphics[width=1.00\linewidth]{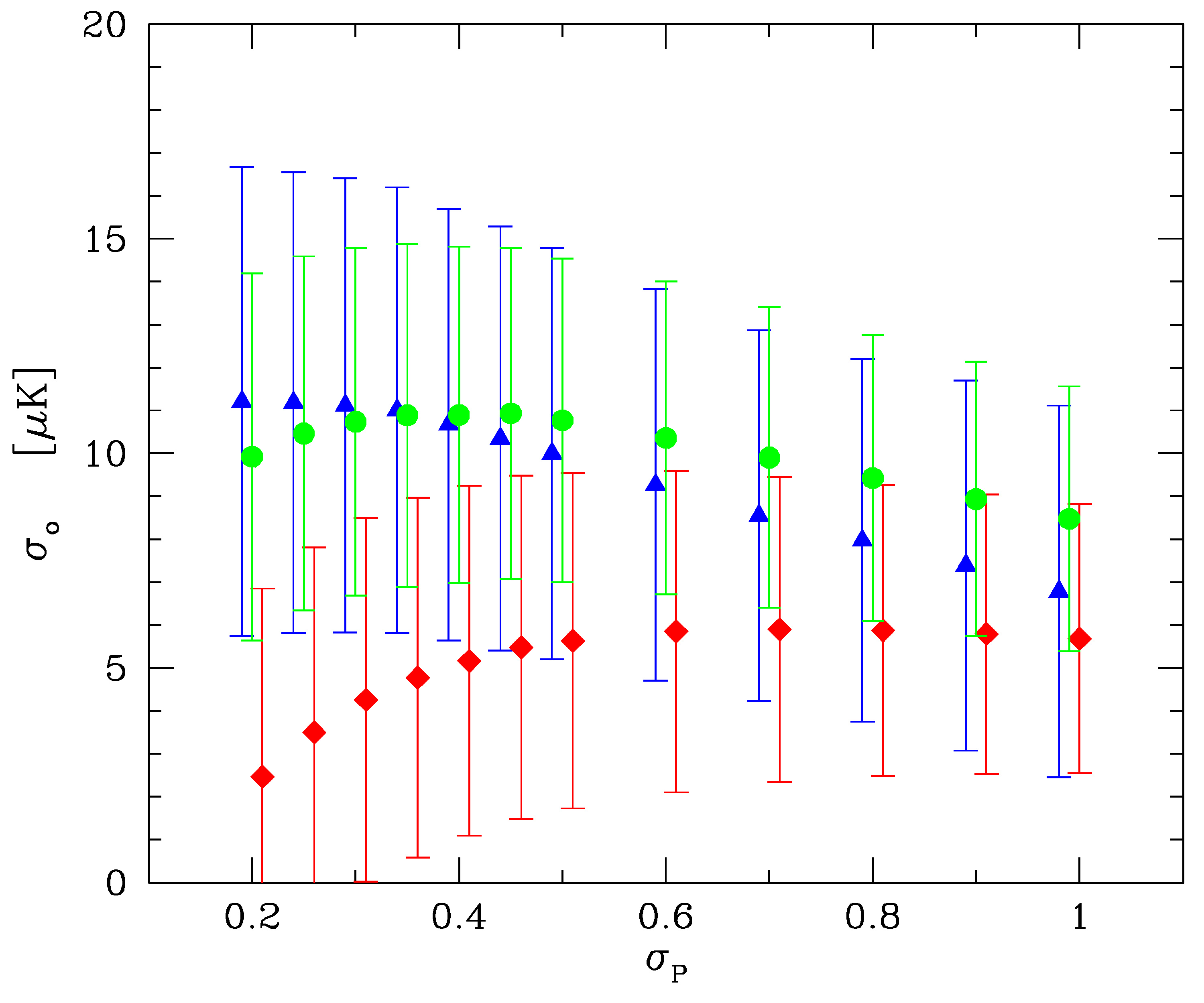}
   \caption{Observational lower limits for the rms fluctuations of the
   ISW signal (Eq.~\ref{eq:sig_o}) for distances $D_{\rm l} - D_{\rm h} =
   1500 - 3000$\,Mpc (triangles), $3000 - 4500$\,Mpc (diamonds),
   and $1500 - 4500$\,Mpc (dots), for a range of filter parameters
   $0.2\,\vartheta \leq \sigma_{\rm P} \leq 1\,\vartheta$
   (Eq.~\ref{eq:filter}),
   for the sphere radius $r_{\rm s}\, =\, 70$\,Mpc.}
   \label{fig:rms_o_flt}
\end{figure}

\ni In Fig.~\ref{fig:rms_o_rs} we plot the $\sigma_{\rm o}$ amplitudes in the
distance bins $1500 - 3000$\,Mpc, $3000 - 4500$\,Mpc and $1500 - 4500$\,Mpc,
for $\sigma_{\rm P} = 0.4\vartheta$ and several sphere radii, $r_{\rm s}$.
The error bars are derived from the MC simulations in a similar way as the rms
uncertainties of coefficient $\rho$. In Table~\ref{tab:one} a selection of the
relevant parameters are listed for sphere radii, $r_{\rm s}$, in the range of
$50 - 110$\,Mpc. The amplitude of $\sgo$ as a function of the  width of
the filter applied to compute the temperature variations, $\sigma_{\rm P}$,
is plotted in Fig.~\ref{fig:rms_o_flt}. We recall that the amplitudes plotted
in figures represent just a fraction of the variance $\sgosq$ that is
`explained' by the model (see the Appendix). Variations of $\sgo$ for different
sphere radii reflect variations of the `unexplained' fraction of $\sgs$ or,
equivalently, an effectiveness of $\DQ$ function to trace the gravitational
potential.

Albeit, the present calculations probe a wide range of linear scales,
a shape of the $\sgo$ variations does not define any particular characteristic
scale of the  gravitational potential fluctuations.
Smooth distributions of $\sgo$ against  $r_{\rm s}$ and
$\sigma_{\rm P}$ indicated in Figs.~\ref{fig:rms_o_rs} and \ref{fig:rms_o_flt}
shows that the $\DQ$ parameter correlates with the potential for a wide range of
sphere radii. At the same time, it  demonstrates
ability of the present method to impose restrictive lower limit on the
amplitude of the ISW signal. One should also notice that the high amplitude of
$\sgo$ is not associated with any exceptional statistical anomaly.

High amplitude of temperature fluctuations generated at the last scattering
surface introduce a strong noise to our $\sgo$ estimates. Therefore
uncertainties of single results are in most cases large, what prevents us from
definite particular conclusions.  Still, some general tendencies are visible.

Both the near and distant samples give positive $\dT - \DQ$ correlations and
provide some constraints on the ISW effect. However, the relationships between
the $\sgo$ and $r_{\rm s}$ in these samples are different. In the near sample
($1500-3000$\,Mpc) the $\sgo$ amplitudes are relatively high at small sphere
radii and diminish with increasing $r_{\rm s}$. The effect is likely caused by
the overall quasar distribution in the sample.  Strong radial density gradient
introduces a bias to quasar counts in spheres, and the effect is dominating for
large $r_{\rm s}$.  The opposite tendency is present in the distant sample
($3000-4500$\,Mpc) with roughly stable quasar density. In such setting the
Poissonian fluctuations are the main source of noise, and their effect on
counts is less pronounced for larger spheres.

\begin{table}
\caption{Interesting parameters on the $\DQ - \dT$ correlations
 for $\sigma_{\rm P} = 0.4 \vartheta$, in distances $1500-3000$\,Mpc and $1500-4500$\,Mpc.}
\label{tab:one}
\begin{tabular}{ccccc}
\hline
{\tiny (1)}   &   {\tiny (2)}     &    {\tiny (3)}     &   {\tiny (4)}     &   {\tiny (5)}      \\
$r_{\rm s}^a$ &     $\rho$        & $\sigma_{\rm o}^b$ &      $\rho$       & $\sigma_{\rm o}^b$ \\
              & \multicolumn{2}{c}{~~$[1500 - 3000]$}  & \multicolumn{2}{c}{~~$[1500 - 4500]$}  \\
\hline
  $~50$       & $0.059 \pm 0.025$ & $12.4 \pm 6.1$     & $0.053 \pm 0.021$ & $10.1 \pm 4.4$     \\
  $~60$       & $0.068 \pm 0.030$ & $12.0 \pm 5.6$     & $0.066 \pm 0.024$ & $11.6 \pm 4.5$     \\
  $~70$       & $0.077 \pm 0.034$ & $10.7 \pm 5.0$     & $0.077 \pm 0.027$ & $10.9 \pm 3.9$     \\
  $~80$       & $0.090 \pm 0.039$ & $10.4 \pm 4.7$     & $0.092 \pm 0.031$ & $11.7 \pm 4.0$     \\
  $~90$       & $0.092 \pm 0.044$ & $~9.6 \pm 4.6$     & $0.099 \pm 0.035$ & $10.9 \pm 3.8$     \\
  $100$       & $0.099 \pm 0.048$ & $~8.9 \pm 4.4$     & $0.108 \pm 0.038$ & $10.6 \pm 3.7$     \\
  $110$       & $0.102 \pm 0.052$ & $~7.8 \pm 4.2$     & $0.116 \pm 0.042$ & $10.0 \pm 3.7$     \\
\hline
\end{tabular}

\small { $a$ - in Mpc, $b$ - in $\umu$K.}
\end{table}


\section{Discussion}
\label{sec:discussion}

The average number of quasars in the SDSS catalogue found within a sphere  of
radius $r_{\rm s}\, =\, 100$\,Mpc varies from $6.6$ at distance $1500$\,Mpc to
$2.2$ at $3000$\,Mpc, and $2.1$ at $4500$\,Mpc. Thus, the Poisson noise
strongly perturbs estimates of the average matter density, particularly at
smaller $r_{\rm s}$ (see also \citealt{nadathur17}). It effectively precludes
attempts to directly estimate space distribution of gravitational potential.
Instead we define the parameter, $\DQ$, determined by the quasar space
distribution, that is statistically correlated with the potential. We
then measure the $\dT - \DQ$ correlation amplitude to determine $\sgo$ -- a
fraction of the temperature fluctuations that can be explained by this
correlation. The signal characterized by $\sgo$ constitutes the lower limit for
total fluctuations generated between $D_{\rm l}$ and $D_{\rm h}$.  In the
calculations, the origin of the correlation is not decided. Albeit, it is
natural to interpret the signal within the framework of the ISW effect.

The amplitude of the CMB rms fluctuations derived from the $\dT - \DQ$
correlation, or the lower limit of the ISW effect, $\sgo$, is determined with
limited accuracy.  For the distant sample a significance of the measurement
does not exceed $2\sigma$, Thus, taken at its face value, the distant sample is
consistent with no ISW signal. However, the near and distant samples combined
together give the significance of the detection higher than each sample
separately.  For $r_{\rm s} = 80$\,Mpc the correlation coefficient $\rho =
0.0924$ differs from zero at $2.9\,\sigma$. The correlations in the near sample
alone exceed $2\,\sigma$ for $r_{\rm s}$ between $40$ and $100$\,Mpc.

In this context, a couple of findings in the present investigation are to some
extent surprising. The $\dT - \DQ$ correlations show {\it a posteriori} that
quasar number counts in spheres provide descriptive assessments on the
gravitational potential along the line of sight, and constrain the
amplitude of the ISW effect.  More numerous sample of objects would allow to
model the space distribution of the potential and to reduce the shot noise.
Presumably, a correlation of such new statistics with $\dT$ would impose still
stronger constraints.

\begin{figure}
   \includegraphics[width=1.00\linewidth]{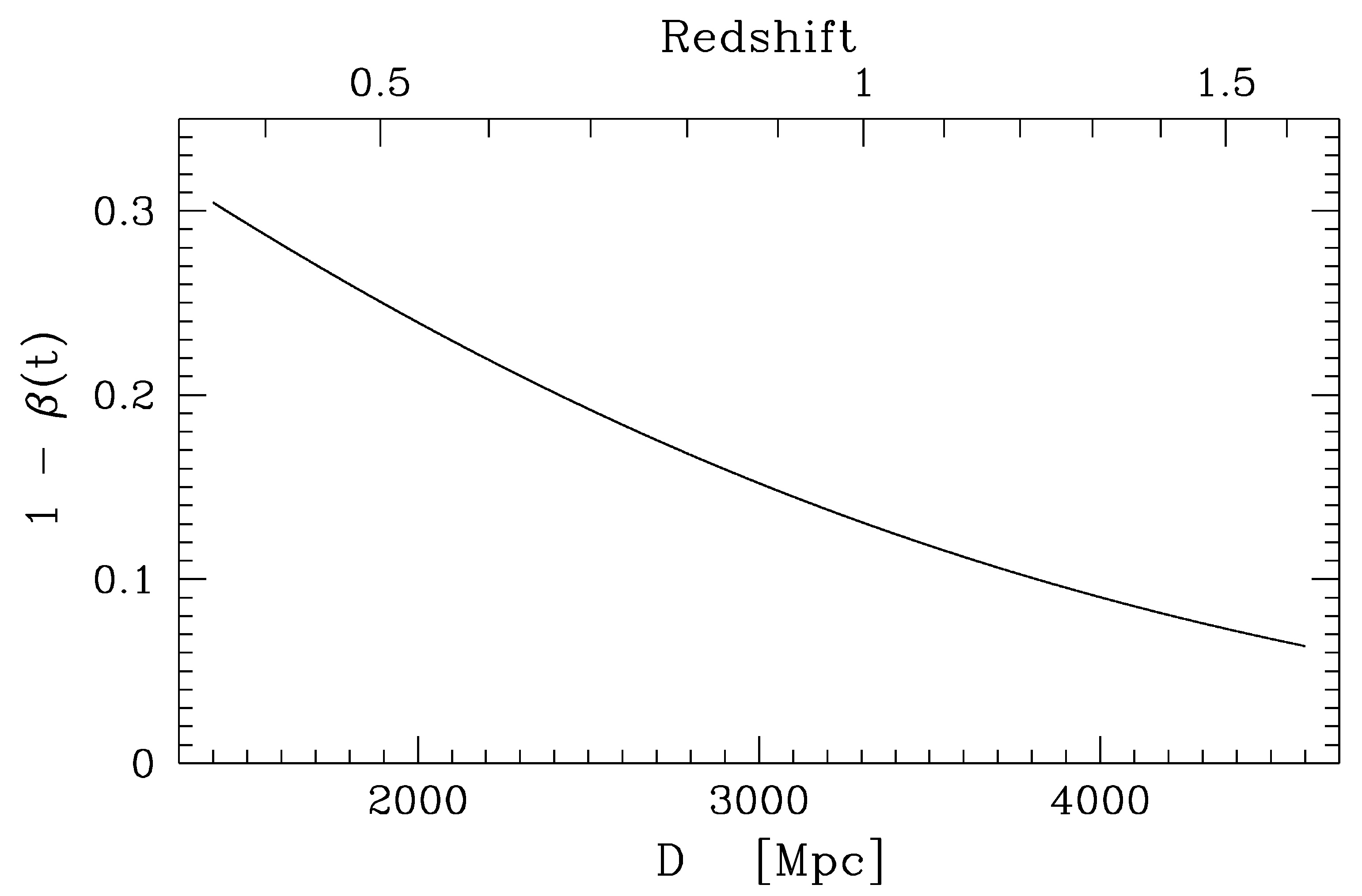}
   \caption{Variations of the $1 - \beta$ parameter that define magnitude
   of the ISW effect (Eq.~\ref{eq:phi_dot}) for $\Omega_\Lambda = 0.70$ and
   $\Omega_{\rm m} = 0.3$.}
   \label{fig:ombeta}
\end{figure}

Magnitude of the ISW effect depends on the parameter $1 - \beta(t)$ in
Eq.~\ref{eq:phi_dot}. For the assumed $\Omega_\Lambda$ and $\Omega_{\rm m}$ in
the $\Lambda$CDM model, $1 - \beta(t)$ rises with decreasing redshift (see
Fig~\ref{fig:ombeta}). Our results qualitatively reflect this tendency.
The most significant estimates of $\sgo$ for the near bin are higher than the
best estimate for the distant bin. However, sizable uncertainties prevent us
from any definite conclusions.

In the paper we examine the correlation between the CMB temperature and
the entire quasar population disregarding any particular configurations of
objects, while a large number of the ISW effect investigations concentrate on
distinct, well-defined galaxy structures, supervoids or superclusters, and
their imprint on the CMB map. Thus, our results and several recent
investigations are not directly comparable, and the objective of this
discussion is limited to qualitative assessments to what extent the present
estimates of the ISW amplitude are consistent with other reports. Nevertheless,
such observation is still instructive because of the persistent controversy
surrounding the positive detection of the ISW signal at a level of several
$\mu$K.  

The present statistics provides only the lower limit for the signal, but it is
free from biases introduced by various void/cluster finding algorithms, and it
is essentially free from the systematic errors introduced by simulations.
Although our method is based on
the simplified relationship between gravitational potential and the density
distribution, it yields restrictive constraints on the ISW effect. Apparently,
our limits at a level of several $\mu$K favour high estimates of the ISW
signal, comparable to those found by \citet{nadathur16} and exceed the best
estimates predicted for the $\Lambda$CDM cosmology (see their Fig.~3). 

Potentially informative is comparison of the present results with the ISW model
maps obtained by \citet{watson14a} based on Jubilee simulations. In their
Fig.~7 $1\sigma$ temperature fluctuations expected in the standard $\Lambda$CDM
cosmology are shown separately for a number of redshift shells of fixed width
$\Delta z = 0.1$.  In five such shells matching our near sample $1500 -
3000$\,Mpc ($0.38 < z < 0.88$), predicted $1\sigma$ fluctuations drop from
$\sim 3\,\mu$K for the nearest bin to $\lesssim 2\,\mu$K for the farthest one.
The joint effect of all five shells depends strongly on the correlations
between the shells. For the perfect correlation the total fluctuation amplitude
generated between $1500$ and $3000$\,Mpc would approach our best estimate of
$\sgo$ at $11 - 12\,\mu$K (see Watson's et al. Fig. 7). Visual inspection
of maps in \citet{watson14a} Fig.~6 shows that the ISW signal is in fact
strongly correlated over the examined redshift range, and  the expected
$1\sigma$ fluctuations are comparable to our $\sgo$.

Statistical significance of our $\sgo$ measurements in the distant bin
$3000-4500$\,Mpc ($0.88 < z < 1.57$) is low (Fig.~\ref{fig:rms_o_rs}). However,
positive $\dT - \DQ$ correlations over a wide range of the test $r_{\rm s}$
values indicate a likely detection of the ISW effect also in this bin. The best
estimates of $\sgo$ reach $\sim 8\,\mu$K, and the \citet{watson14a}
signal\footnote{The \citet{watson14a} results extend only to $z \approx 1.4$,
but the expected contribution from  $1,4 < z < 1.57$ is minute.} has comparable
amplitude assuming significant correlations of fluctuations in the
contributing redshift shells.

Apparent coincidences between the \citet{watson14a} amplitudes and our $\sgo$
estimates is puzzling because both results represent different quantities.  The
total rms fluctuations predicted from the simulations correspond to our $\sgs$
in Eq.~\ref{eq:sigmas}, and not to $\sgo$ alone. Clearly, our statistics based
on quasar counts along the line of sight is unable to reconstruct variations of
the gravitational potential in a way it is done by the present day cosmological
simulations. Nevertheless, this statistics provides constraining limit on the
ISW effect.  To examine if in fact the observed signal is reproduced in the
$\Lambda$CDM cosmological simulations, one should use the simulations to
generate the model ISW imprint in the CMB maps, and to create quasar
distribution analogous to the SDSS catalogue. Then, one should apply the $\DQ$
statistics to the mock quasar data to obtain constraints on the ISW effect.
Eventually the comparison of these constraints with the model ISW would allow us
to answer if the observed ISW signal is compatible with the $\Lambda$CDM
cosmology.

Possible origins for the observed amplitude of the ISW effect are
discussed by \citet{cai17}, who investigated both lensing and temperature
signatures of voids on the CMB. Using the DR12 SDSS CMASS galaxy sample they
confirm reports that the temperature variations associated with large voids is
substantially stronger than those predicted for the $\Lambda$CDM model, but
'the amplitude of the lensing convergence signal $\Delta \kappa$ is a very good
match to $\Lambda$CDM'.  \citet{cai17} indicate that both observables `may
provide valuable information on cosmology and gravity'.

Stronger than expected correlation of the CMB temperature with the matter
density fluctuations could indicate some departures from the standard
$\Lambda$CDM model. This point was raised by \citet{nadathur12} and
\citet{kovacs17}. In particular, such inconsistency could imply some dark
matter - dark energy interactions \citep{olivares08}. However, it is possible
that the observed correlation cannot be attributed solely to the ISW effect,
i.e. some fraction of the signal is produced in a different mechanism. To
support this conjecture, we note that a positive correlation between the
temperature amplitude and quasar (or galaxy) concentration independent from the
ISW effect should also be considered.  Exact contribution of numerous classes
of foreground objects at microwave frequencies is difficult to assess. Simple
interpolation of the `spectrum of the Universe' \citep{hill18} between two
spectral domains adjacent to the CMB maximum, i.e.  radio below $\sim 3\cdot
10^8$\,Hz and far infra red above $\sim 10^{12}$\,Hz, gives in the microwave
range flux exceeding $0.001$ of the relic CMB. This figure surpasses the
potential ISW signal by more than two orders of magnitude.

Despite a `tremendous effort' \citep{kovacs17} put to eliminate foreground
emission, it is possible that some residual contamination of the relic
radiation by sources with spectral energy distribution resembling the CMB
cannot be ruled out.  Such sources would not be removed in the {\it Planck}
data processing aimed to cut-off the  foreground emission. Also, their imprint
on the CMB would not violate the \citet{planckXIX} conclusion that the signal
correlated with the large scale structures is achromatic over a {\it Planck}
spectral range.  Thus, even minute contribution of those hypothetical sources
could remove the discrepancy. One should note, that both average galaxy spectrum
and spectra of various AGN types differ significantly from the CMB spectrum.
Nevertheless, if a small population of those specific sources exists, a
precise calibration of the ISW effects and still more accurate assessment of
all the cosmic background components are needed before potential deviations
from the standard $\Lambda$CDM cosmology are contemplated.


\section*{acknowledgements}
I thank the anonymous reviewer for critical comments 
that helped me to improve the material content of the paper.

\bibliography{soltan_sh}

\appendix

\section[]{Linear correlation statistics}
\label{app:corr}

Here we derive basic relationships between the interesting observables
under the assumption that relevant quantities are linearly correlated.
This concerns in particular the fluctuations 
of the CMB temperature, $\dT$ and the parameter derived from the
quasar space distribution, $\DQ$ in Sec.~\ref{sec:results}.

In the subsequent derivation $t$ and $q$ denote  single measurements of $\dT$
and $\DQ$, respectively. The observed amplitude $t = s + b$, where $s$ is the
temperature fluctuation of the ISW effect generated by the potential
distribution within the considered distance range, $D_{\rm l}$ and $D_{\rm h}$,
and $b$ represents the total `background' fluctuations uncorrelated with the
matter distribution in this area. It is assumed that $b$ and $s$ are
uncorrelated, and the expected values of both variables are equal to zero,
$\mu_{\rm b} = 0$ and $\mu_{\rm s} = 0$. No other specific constraints are
imposed on the distributions of both variables. In particular, the dispersions
$\sgb$ and $\sgs$ are unknown, and satisfy usual relationship
$\sgbsq\,+\,\sgssq\,=\,\sgtsq$, where $\sgt$ denote the observed rms scatter of
$t$.

The variable $q$ by its very nature is correlated with the matter density along
the line of sight. The expected value of the variable, $\mu_{\rm q} = 0$.  It
is also subject to random fluctuations caused by a discrete nature of quasar
data.  Consequently, it is assumed that $q$ is adequately represented by a sum
of two components: $q\,=\,c\,+\,r$, where $c$ represents  quasar `clustering',
i.e. the component actually correlated with the ISW signal, and $r$ describes
random fluctuations produced by Poisson noise. 

In Sec.~\ref{sec:results} statistical characteristics of the $q$ and $t$
distributions are discussed. Their  variances, $\sgtsq$ and $\sgqsq$, are
determined.  It is shown that variables $q$ and $t$ are linearly correlated The
interesting parameters of the distribution of the $r$ component, i.e.  the
expected value  and variance, $\mu_{\rm r}$ and $\sgrsq$ were determined using
the Monte Carlo simulations for all the combinations of sphere radii and
distance bins.  In the whole range of sphere radii considered in the paper the
random scatter of $q$ resulting from discrete quasar distribution constitutes a
dominating fraction of $\sgq$. Consequently, the dispersion of $c$ is small as
compared to $\sgr$, and it is legitimate to assume that the $c$ and $r$ are
uncorrelated, and to put $\sgqsq\,=\,\sgcsq\,+\,\sgrsq$.

Variables $s$ and $c$ are not directly observable. However, amplitude of their
intrinsic correlation generates the correlation between $t$ and $q$ that is
determined from the data.  Therefore, the amplitude of the $q - t$ correlation
allows to estimate some interesting parameters of the joint distribution of
variables $s$ and $c$.  Similarly to the pair of variables $(s,\,c)$, the pair
$(s,\,q)$ is linearly correlated with the relevant parameters: $\mu_s\,=\,0$,
$\mu_q\,=\,0$, $\sgs$, $\sgq$, and $\rho_{sq}$, where only $\sgq$ is directly
determinable from the data.

By means of elementary calculations one gets:
\begin{equation}
\rho_{sq}\, =\, \rho_{sc}\, \frac{\sgc}{\sgq}\,,
\label{eq:a1}
\end{equation}

\begin{equation}
\rho_{sq}\, =\, \beta_{s|q}\, \frac{\sgq}{\sgs}\,, \hspace{15mm}
\rho_{sc}\, =\, \beta_{s|c}\, \frac{\sgc}{\sgs}\,,
\label{eq:a2}
\end{equation}

\ni where $\beta_{s|q}$ and $\beta_{s|c}$  are slopes of the least squares
regression lines of $s$ on $q$ and $s$ on $c$, respectively. Combining
Eqs.~\ref{eq:a1} and \ref{eq:a2}:

\begin{equation}
\frac{\beta_{s|c}}{\beta_{s|q}}\, =\, \frac{\sgqsq}{\sgcsq}\,.
\end{equation}

\ni The variance  of $s$, $\sgssq$, may be decomposed into:

\begin{equation}
\sgssq\, =\, \sigma_{s|c}^2\, +\, \beta_{s|c}^2\, \sgcsq\,,
\end{equation}

\ni where $\sigma_{s|c}$ is the rms scatter of $s$ around the regression
line of $s$ on $c$. Since $\beta_{s|q}\,=\,\beta_{t|q}$ and
$\sgcsq\,=\,\sgqsq\,-\,\sgrsq$\,:

\begin{equation}
\sgssq\, =\, \sigma_{s|c}^2\, +\,
 \beta_{t|q}^2\, \frac{\sgq^4}{\sgqsq\,-\,\sgrsq}\,,
\end{equation}

\ni or

\begin{equation}
\sgssq\, =\,  \sigma_{s|c}^2\, +\, \sgosq\,,
\label{eq:sigmas}
\end{equation}

\ni where

\begin{equation}
\sgo\, =\, \rho_{\rm tq}\,\frac{\sgt\,\sgq}{\sqrt{\sgqsq\,-\,\sgrsq}}\,,
\label{eq:sig_o}
\end{equation}

\ni and we put $\beta_{t|q}\,=\,\rho_{\rm tq}\,\frac{\sgt}{\sgq}$.

In statistics the second term in the right-hand side of Eq.~\ref{eq:sigmas},
$\sgosq$, represents the `explained' or `model' fraction of the total
variance, $\sgssq$, while the first term remains `unexplained' by the model. In
the present investigation, fluctuations of the gravitational potential that
generate the ISW effect are represented by the  $\DQ$ parameter
(Sec.~\ref{sec:results}).  A question how accurate is this approximation is not
addressed in the paper.  Therefore, amplitude of the `unexplained' variance
$\sigma_{s|c}^2$ remains undetermined.

\label{lastpage}

\end{document}